\documentclass[conference]{IEEEtran}
\IEEEoverridecommandlockouts
\usepackage{utfsym}
\usepackage{filecontents}
\usepackage{amsmath,amsfonts}
\usepackage{graphicx}
\usepackage{textcomp}
\usepackage{xcolor}
\usepackage{epsfig,subcaption}
\usepackage{tikz}
\usepackage{bm}
\usepackage{wasysym}
\usepackage{algorithm}
\usepackage{algpseudocode}
\usepackage{diagbox}
\usepackage{makecell}
\usepackage{booktabs}
\usepackage{adjustbox}

\usepackage{threeparttable}
\usepackage{latexsym,fancyhdr,hyperref}
\usepackage{adjustbox,multirow, booktabs, hhline}
\usepackage{authblk}
\let\labelindent\relax
\usepackage{enumitem}
\setenumerate[1]{itemsep=0pt,partopsep=0pt,parsep=\parskip,topsep=5pt,leftmargin=10pt}
\setitemize[1]{itemsep=0pt,partopsep=0pt,parsep=\parskip,topsep=5pt,leftmargin=10pt}
\setdescription{itemsep=0pt,partopsep=0pt,parsep=\parskip,topsep=5pt,leftmargin=10pt}

\def\hlb{\color{black}}

\def\BibTeX{{\rm B\kern-.05em{\sc i\kern-.025em b}\kern-.08em
    T\kern-.1667em\lower.7ex\hbox{E}\kern-.125emX}}

\makeatletter

\newcommand{\linebreakand}{%
  \end{@IEEEauthorhalign}
  \hfill\mbox{}\par
  \mbox{}\hfill\begin{@IEEEauthorhalign}
}
\makeatother

\begin{document}


\date{}

\title{ \small{Accepted by the 31st Network and Distributed System Security Symposium (NDSS 2024)}
\\
\Huge{Parrot-Trained Adversarial Examples: Pushing the Practicality of Black-Box Audio Attacks against Speaker Recognition Models}}


\author{
    {Rui Duan}\\
    University of South Florida\\
    Tampa, USA\\
    ruiduan@usf.edu
    \and
    {Zhe Qu}\\
    Central South University\\
    Changsha, China\\
    zhe\_qu@csu.edu.cn
    \and
    {Leah Ding}\\
    American University\\
    Washington, DC, USA\\
    ding@american.edu
    \linebreakand   
    {Yao Liu}\\
    University of South Florida\\
    Tampa, USA\\
    yliu@cse.usf.edu
    \and
    
    {Zhuo Lu}\\
    University of South Florida\\
    Tampa, USA\\
    zhuolu@usf.edu
}

\maketitle

\begin{abstract}
Audio adversarial examples (AEs) have posed significant security challenges to real-world speaker recognition systems. Most black-box attacks still require certain information from the speaker recognition model to be effective (e.g., keeping probing and requiring the knowledge of similarity scores). This work aims to push the practicality of the black-box attacks by minimizing the attacker's knowledge about a target speaker recognition model. Although it is not feasible for an attacker to succeed with completely zero knowledge, we assume that the attacker only knows a short (or a few seconds) speech sample of a target speaker. Without any probing to gain further knowledge about the target model, we propose a new mechanism, called parrot training, to generate AEs against the target model. Motivated by recent advancements in voice conversion {\hlb(VC)}, we propose to use the one short sentence knowledge to generate more synthetic speech samples that sound like the target speaker, called parrot speech. Then, we use these parrot speech samples to train a parrot-trained (PT) surrogate model for the attacker. Under a joint transferability and perception framework, we investigate different ways to generate AEs on the PT model (called PT-AEs) to ensure the PT-AEs can be generated with high transferability to a black-box target model with good human perceptual quality.
Real-world experiments show that the resultant PT-AEs achieve the attack success rates of $45.8\%$--$80.8\%$ against the open-source models in the digital-line scenario and $47.9\%$--$58.3\%$ against smart devices, including Apple HomePod (Siri), Amazon Echo, and Google Home, in the over-the-air scenario.\footnote{Our attack demo can be found at: \href{https://sites.google.com/view/pt-attack-demo}{https://sites.google.com/view/pt-attack-demo} }

\end{abstract}

\section{Introduction}
Adversarial speech attacks against speech recognition \cite{carlini2018audio, yuan2018commandersong, li2020advpulse, taori2019targeted, wang2020towards, chen2020devil, du2020sirenattack, zheng2021black} and speaker recognition \cite{du2020sirenattack, chen2019real, zheng2021black}
have become one of the most active research areas of machine learning in computer audio security. These attacks craft audio adversarial examples (AEs) that can spoof the speech classifier in either white-box \cite{carlini2018audio, yuan2018commandersong,li2020advpulse,guo2022specpatch} or black-box settings \cite{wang2020towards, chen2020devil, du2020sirenattack, zheng2021black, chen2019real,liu2022evil,abdullah2019hear}. Compared with white-box attacks that require the full knowledge of a target audio classification model, black-box attacks do not assume the full knowledge and have been investigated in the literature under different attack scenarios \cite{chen2019real,zheng2021black}. Despite the substantial progress in designing black-box attacks, they can still be challenging to launch in real-world scenarios in that the attacker is still required to gain information from the target model.

Generally, the attacker can use a query (or probing) process to gradually know the target model: repeatedly sending a speech signal to the target model, then measuring either the confidence level/prediction score \cite{chen2020devil, du2020sirenattack, chen2019real} or the final output results \cite{zheng2021black,yusmack} of a classifier. The probing process usually requires a large number of interactions (e.g., over 1000 queries \cite{yusmack}), which can cost substantial labor and time. This may work in the digital line, such as interacting with local machine learning models (e.g., Kaldi toolkit \cite{povey2011kaldi}) or online commercial platforms (e.g., Microsoft Azure \cite{Microsoft_Azure}). However, it can be even more cumbersome, if not possible, to probe physical devices because today’s smart devices (e.g., Amazon Echo \cite{test_amazon}) accept human speech over the air. Moreover, some internal knowledge of the target model still has to be assumed known to the attacker (e.g., the access to the similarity scores of the target model \cite{chen2019real,yusmack}). Two recent studies further limited the attacker's knowledge to be (i) \cite{zheng2021black} only knowing the target speaker's one-sentence speech \cite{zheng2021black} and requiring probing to get the target model's hard-label (accept or reject) results (e.g., over 10,000 times) and (ii) \cite{chen2023qfa2sr} only knowing one-sentence speech for each speaker enrolled in the target model.

In this paper, we present a new, even more practical perspective for black-box attacks against speaker recognition. We first note that the most practical attack assumption is to let the attacker know nothing about the target model and never probe the model. However, such completely zero knowledge for the attacker unlikely leads to effective audio AEs. We have to assume some knowledge but keep it at the minimum level towards the attack practicality. Our work limits the attacker's knowledge to be only a one-sentence (or a few seconds) speech sample of her target speaker without knowing any other information about the target model. The attacker has neither knowledge of nor access to the internals of the target model. Moreover, she does not probe the classifier and needs no observation of the classification results (either soft or hard labels). To the best of our knowledge, our assumption of the attacker's knowledge is the most restricted compared with prior work (in particular with the two recent attacks \cite{zheng2021black, chen2023qfa2sr}).

Centered around this one-sentence knowledge of the target speaker, our basic attack framework is to (i) propose a new training procedure, called parrot training, which generates a sufficient number of synthetic speech samples of the target speaker and uses them to construct a parrot-trained (PT) model for a further transfer attack, and (ii) systematically evaluate the transferability and perception of different AE generation mechanisms and create PT-model based AEs (PT-AEs) towards high attack success rates and good audio quality.

Our motivation behind parrot training is that the recent advancements in the {\hlb voice conversion (VC)}
domain have shown that the one-shot speech methods \cite{chou2019one,lu2019one,wu2020one,chen2021again} are able to leverage the semantic human speech features to generate speech samples that sound like a target speaker's voice in different linguistic contents. Based on the attacker's one-sentence knowledge, we should be able to generate different synthetic speech samples of her target speaker and use them to build a PT model for speaker recognition. Our feasibility evaluations show that a PT model can perform similarly to a ground-truth trained (GT) model that uses the target speaker's actual speech samples.

The similarity between PT and GT models creates a new, interesting question of transferability: if we create a PT-AE from a PT model, can it perform similarly to an AE generated from the GT model (GT-AE) and transfer to a black-box target GT model? Transferability in adversarial machine learning is already an intriguing concept. It has been observed that the transferability depends on many aspects, such as model architecture, model parameters, training dataset, and attacking algorithms \cite{mao2022transfer,liu2016delving}. Existing AE evaluations have been primarily focused on GT-AEs on GT models without involving synthetic data. As a result, we conduct a comprehensive study on PT-AEs in terms of their generation and quality.

\begin{itemize}
\item Generation: As an audio AE consists of the original signal and a perturbation signal. One essential difference in existing studies lies in finding the perturbation signal from different types of audio waveforms, which we call {\it carriers} in this paper. In particular, we summarize the carriers into the following major types: (i) noise carriers, which are the results of traditional methods \cite{chen2019real,zheng2021black} during their search for the perturbation signals in the unrestricted $L_p$ space. (ii) feature-twisted carriers that are perturbation signals generated by only varying the auditory features of the original signal itself \cite{yusmack, duan2022perception, abdullah2019hear,chen2023qfa2sr}, (iii) environmental sound carriers that are produced by environmental sounds \cite{deng2022fencesitter}. Based on the built PT model, we create and evaluate PT-AEs based on these three types of carriers.

\item Quality: We first need to define a quality metric to quantify whether a PT-AE is good or not. There are two important factors of PT-AEs: (i) transferability of PT-AEs to a black-box target model. We adopt the match rate, which has been comprehensively studied in the image domain \cite{mao2022transfer}, to measure the transferability. The match rate is defined as the percentage of PT-AEs that can still be misclassified as the same target label on a black-box GT model. (ii) The perception quality of audio AEs. We conduct a human study to let human participants rate the speech quality of AEs with different types of carriers in a unified scale of perception score from 1 (the worst) to 7 (the best) commonly used in speech evaluation studies \cite{gemmeke2017audio,11wester,bunton2007listener,anand2019objective,patel2010perceptual,darley1969differential}, and then build regression models to predict human scores of speech quality. However, these two factors are generally contradictory, as a high level of transferability likely results in poor perception quality. We then define a new metric called {\it transferability-perception ratio (TPR)} for PT-AEs generated using a specific type of carriers. This metric is based on their match rate and average perception score, and it quantifies the level of transferability a carrier type can achieve in degrading a unit score of human perception. A high TPR can be interpreted as high transferability achieved by a relatively small cost of perception degradation.
\end{itemize}

\begin{table}[t]
    \centering
    \caption{Summary of common attack strategies.}\label{tab:audio_attacks}
    \vspace{-0.2cm}
    \begin{adjustbox}{width=0.45\textwidth}
    \begin{threeparttable}
    \begin{tabular}{ccccc}
    \hline
    Attack Strategy         & \begin{tabular}[c]{@{}l@{}} Attack\\Scenario \end{tabular} & \begin{tabular}[c]{@{}l@{}}Queries\\Needed \end{tabular}       & \begin{tabular}[c]{@{}l@{}}Knowledge\\Required\end{tabular} & \begin{tabular}[c]{@{}c@{}}Human\\Perception\end{tabular} \\ \hline
    Carlini et al.\cite{carlini2018audio}  & \multicolumn{1}{c}{White-box}                             & $\sim$1000  &gradient info  & \usym{2717}                                         \\ 
    CommanderSong\cite{yuan2018commandersong}   & \multicolumn{1}{c}{White-box}                             & $\sim$100  &gradient info   & \usym{2717}                                            \\ 
    Psychoacoustic\cite{qin2019imperceptible}  & \multicolumn{1}{c}{White-box}                             & $\sim$5000  &gradient info  & \usym{2713}                                            \\ 
    AdvPulse\cite{li2020advpulse}       & \multicolumn{1}{c}{White-box}                             & $\sim$2000   &gradient info & \usym{2717}                                                                                 \\ 
    SpecPatch\cite{guo2022specpatch}      & \multicolumn{1}{c}{White-box}                             & $\sim$1000  &gradient info  & \usym{2713}                                                                                     \\ 
    Taori et al.\cite{taori2019targeted}    & Black-box                                                 & $\sim$300,000  &soft label   & \usym{2717}                                             \\ 
    SGEA\cite{wang2020towards}            & Black-box                                                 & $\sim$300,000 &soft label & \usym{2717}                                            \\ 
    Devil's Whisper\cite{chen2020devil} & Black-box                                                 & $\sim$1500  &soft label  & \usym{2717}                                           \\ 
    FakeBob\cite{chen2019real}         & Black-box                                                 & $\sim$5000  &soft label  & \usym{2717}                                            \\ 
    OCCAM\cite{zheng2021black}           & Black-box                                                 & $\sim$10,000 &hard label & \usym{2717}                                           \\ 
    TAINT\cite{liu2022evil}           & Black-box                                                 & $\sim$1500 &hard label   & \usym{2713}                                        \\ 
    SMACK\cite{yusmack}            & Black-box                                                 & $\sim$1000  &soft label  & \usym{2713}                                         \\ 
    
    QFA2SR \cite{chen2023qfa2sr}            & Black-box                                                 & 0         & each speaker's sample
        &\usym{2717}                                          \\ 
        PT-AE attack            & Black-box                                                 & 0         & target speaker's sample &\usym{2713}                                                                              \\ \hline
    \end{tabular}
    \begin{tablenotes}
            \footnotesize
            \item  (i) Queries: indicating the typical number of probes need to interact with the black-box target model. (ii) Soft level: the confidence score \cite{chen2020devil} or prediction score \cite{taori2019targeted,wang2020towards,chen2020devil,chen2019real,yusmack} from the target model. (iii) Hard label: accept or reject result \cite{zheng2021black,liu2022evil} from the target model. (iv) QFA2SR \cite{chen2023qfa2sr} requires the speech sample of each enrolled speaker in the target model. (v) Human perception means integrating the human perception factor into the AE generation.
          \end{tablenotes}
        \end{threeparttable}
    \end{adjustbox}    
    \vspace{-0.4cm}
    \end{table}

Under the TPR framework, we formulate a two-stage PT-AE attack that can be launched over the air against a black-box target model. In the first stage, we narrow down from a full set of carriers to a subset of candidates with high TPRs for the attacker's target speaker. In the second stage, we adopt an ensemble learning-based formulation \cite{liu2016delving} that selects the best carrier candidates from the first stage and manipulates their auditory features to minimize a joint loss objective of attack effectiveness and human perception. Real-world experiments show that the proposed PT-AE attack achieves the success rates of 45.8\%--80.8\% against open-source models in the digital-line scenario and 47.9\%--58.3\% against smart devices, including Apple HomePod (Siri), Amazon Echo, and Google Home, in the over-the-air scenario. Compared with two recent attack strategies Smack \cite{yusmack} and QFA2SR \cite{chen2023qfa2sr}, our strategy achieves improvements of 263.7\% (attack success) and 10.7\% (human perception score) over Smack, and 95.9\% (attack success) and 44.9\% (human perception score) over QFA2SR. Table~\ref{tab:audio_attacks} provides a comparison of the required knowledge between the proposed PT-AE attack and existing strategies. 

Our major contribution can be summarized as follows. (i)  We propose a new concept of the PT model and investigate state-of-the-art VC methods to generate parrot speech samples to build a surrogate model for an attacker with the knowledge of only one sentence speech of the target speaker. (ii) We propose a new TPR framework to jointly evaluate the transferability and perceptual quality for PT-AE generations with different types of carriers. (iii) We create a two-stage PT-AE attack strategy that has been shown to be more effective than existing attacks strategies, while requiring the minimum level of the attacker's knowledge.

\section{Background and Motivation}\label{Sec:background}
In this section, we first introduce the background of speaker recognition, then describe black-box adversarial attack formulations to create audio AEs against speaker recognition.

\subsection{Speaker Recognition}
Speaker recognition becomes more and more popular in recent years. It brings machines the ability to identify a speaker via his/her personal speech characteristics, which can provide personalized services such as convenient login \cite{Fidelity} and personalized experience \cite{Alexa} for calling and messaging. 
{\hlb 
Commonly, the speaker recognition task includes three phases: training, enrollment, and recognition. It is important to highlight that speaker recognition tasks [29], [118], [113] can be either (i) multiple-speaker-based speaker identification (SI) or (ii) single-speaker-based speaker verification (SV). Specifically, SI can be divided into close-set identification (CSI) and open-set identification (OSI) \cite{deng2022fencesitter, chen2019real}. We provide detailed information in Appendix~\ref{sec:SR}.}

\subsection{Adversarial Speech Attacks}
Given a speaker recognition function $f$, which takes an input of the original speech signal $x$ and outputs a speaker's label $y$, an adversarial attacker aims to find a small perturbation signal $\delta \in \Omega$ to create an audio AE $x+\delta$ such that
\begin{equation}\label{Eq:BasicObj}
\quad f(x+\delta) = y_t, \quad D(x,x+\delta) \leq \epsilon,
\end{equation}
where $y_t \neq y$ is the attacker's target label; $\Omega$ is the search space for $\delta$; $D(x,x+\delta)$ is a distance function that measures the difference between the original speech $x$ and the perturbed speech $x+\delta$ and can be the $L_p$ norm based distance \cite{chen2019real, zheng2021black} or a measure of auditory feature difference (e.g., qDev \cite{duan2022perception} and NISQA \cite{yusmack}); and $\epsilon$ limits the change from $x$ to $x+\delta$. 

A common white-box attack formulation \cite{carlini2018audio,li2020advpulse} to solve \eqref{Eq:BasicObj} can be written as
\begin{eqnarray}
  \arg\min_{\delta \in \Omega} \mathcal{J}(x+\delta,y_t) + c\,D(x,x+\delta)\label{Eq:BasicWhiteBox},
\end{eqnarray}
where $\mathcal{J}(\cdot,\cdot)$ is the prediction loss in the classifier $f$ when associating the input $x+\delta$ to the target label $y_t$, which is assumed to be known by the attacker; and $c$ is a factor to balance attack effectiveness and change of the original speech.

A black-box attack has no knowledge of $\mathcal{J}(\cdot,\cdot)$ in \eqref{Eq:BasicWhiteBox} and thus has to adopt a different type of formulation depending on what other information it can obtain from the classifier $f$. If the attack can probe the classifier that gives a binary (accept or reject) result, the attack \cite{zheng2021black, liu2022evil} can be formulated as
\begin{equation}\label{Eq:ProbBlackBox}
    \arg\min_{\delta\in\Omega} \, \mathcal{L}(x+\delta) \!=\! \left\{
    \begin{array}{ll}
        \!\!D(x,x+\delta)  & \quad\text{if}~f(x+\delta) \!=\! y_t , \\
        \!\!+\infty & \quad \text{otherwise}.
    \end{array}
\right.
\end{equation}
Since \eqref{Eq:ProbBlackBox} contains $f(x+\delta)$, the attacker has to create a probing strategy to continuously generate a different version of $\delta$ and measure the result of $f(x+\delta)$ until it succeeds. Accordingly, a large number of probes (e.g., over 10,000 \cite{zheng2021black}) are required, which makes real-world attacks less practical against commercial speaker recognition models that accept speech signals over the air.

\subsection{Design Motivation}
To overcome the cumbersome probing process of a black-box attack, we aim to find an alternative way to create practical black-box attacks. Given the fact that a black-box attack is not possible without probing or knowing any knowledge of a classifier, we adopt an assumption of prior knowledge used in \cite{zheng2021black} that the attacker possesses a very short audio sample of the target speaker (note that \cite{zheng2021black} has to probe the target model in addition to this knowledge). This assumption is more practical than letting the attacker know the classifier's internals. Given this limited knowledge, we aim to remove the probing process and create effective AEs.

To this end, we go back to the white-box attack formulation in \eqref{Eq:BasicWhiteBox} and try to build a local function $\mathcal J^*$ similar to the loss prediction function $\mathcal J$ in \eqref{Eq:BasicWhiteBox}, then replace $\mathcal J$ with $\mathcal J^*$ to create an audio AE. This may look like a traditional transfer attack strategy \cite{chen2020devil}. But the key difference is that the traditional transfer attack still needs to keep probing the classifier (e.g., 1500 queries \cite{chen2020devil}) to build the local model $\mathcal J^*$; in contrast, the attacker here only has a very short sample of the target speaker to construct $\mathcal J^*$ without probing.

As a result, the first challenge we need to solve is how to build $\mathcal J^*$ based on a very short audio sample. As human speech is semantic, the recent advancements in the {\hlb VC} domain have shown that the one-shot speech methods \cite{chou2019one,lu2019one,wu2020one,chen2021again}, commonly taking a source speaker's audio sample and a target speaker's sample as two inputs, are able to output a speech sample that sounds like the target speaker's voice in the source speaker's linguistic content. Hence, we are motivated to explore the feasibility of using the one-shot speech methods to create synthetic audio data of the attacker's target speaker. As this process is similar to training a parrot to reproduce more speech samples that can mimic the target speaker, we call them {\em parrot speech samples}, based on which we train the local model $\mathcal J^*$ to create audio AEs. We call this method {\em parrot training}, in contrast to the {\em ground-truth training} that uses a speaker's real audio samples to train. 


\begin{figure}[t]
    \centering
    \includegraphics[width=0.48\textwidth]{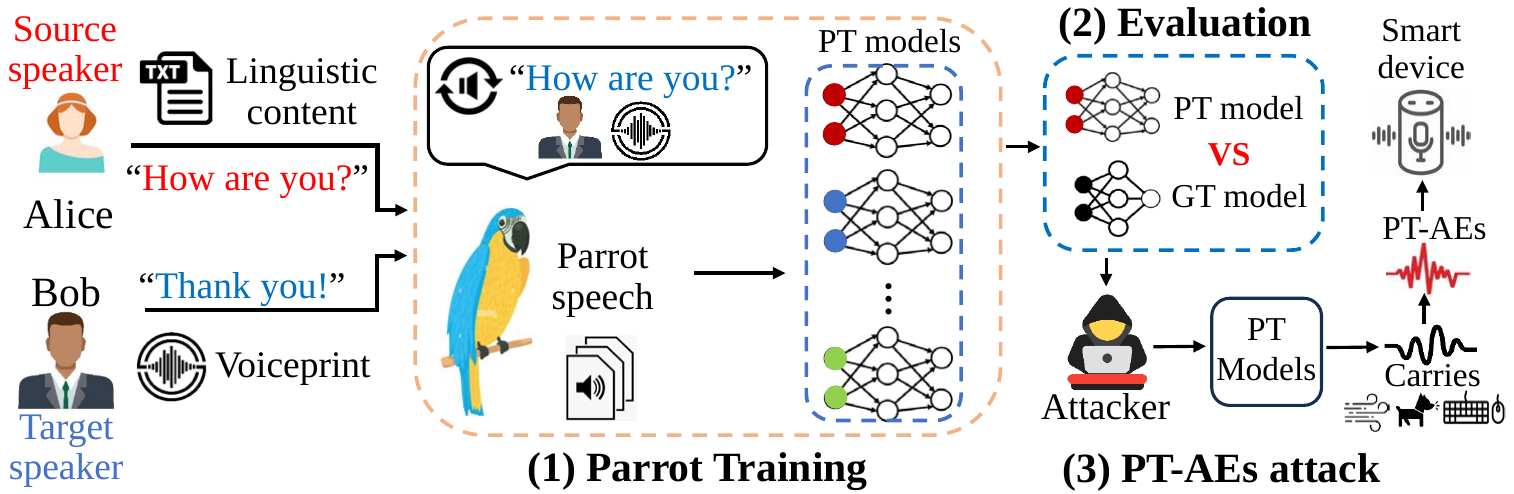}
    \caption{The procedure of parrot training based black-box attack.}
    \label{Fig:parrot_training}
    \vspace{-0.4cm}
\end{figure}

Existing studies have focused on a wide range of aspects regarding ground-truth trained AEs (GT-AEs). The concepts of parrot speech and parrot training create a new type of AEs, parrot-trained AEs (PT-AEs), and also raise three major questions of the feasibility and effectiveness of PT-AEs towards a practical black-box attack: (i) Can a PT model approximate a GT model? (ii) Are PT-AEs built upon a PT model as transferable as GT-AEs against a black-box GT model? (iii) How to optimize the generation of PT-AEs towards an effective black-box attack? Fig.~\ref{Fig:parrot_training} shows the overall procedure for us to address these questions towards a new, practical and non-probing black-box attack: (1) we propose a two-step one-shot conversion method to create parrot speech for parrot training in Section~\ref{Sec:parrot}; (2) we study different types of PT-AE generations from a PT model regarding their transferability and perception quality in Section~\ref{Sec:perception}; and (3) we formulate an optimized black-box attack based on PT-AEs in Section~\ref{Sec:ensemble}. Then, we perform comprehensive evaluations to understand the impact of the proposed attack on commercial audio systems in Section~\ref{Sec:evaluation}.

\subsection{Threat Model}

In this paper, we consider an attacker that attempts to create an audio AE to fool a speaker recognition model such that the model recognizes the AE as a target speaker's voice. We adopt a black-box attack assumption that the attacker has no knowledge about the architecture, parameters, and training data used in the speech recognition model. We assume that the attacker has a very short speech sample (a few seconds in our evaluations) of the target speaker, which can be collected in public settings \cite{zheng2021black}, but the sample is not necessarily used for training in the target model. We focus on a more realistic scenario where the attacker does not probe the model, which is different from most black-box attack studies \cite{yusmack,chen2019real, zheng2021black} that require many probes. We assume that the attacker needs to launch the over-the-air injection against the model (e.g., Amazon Echo, Apple HomePod, and Google Assistant).

\section{Parrot Training: Feasibility and Evaluation}\label{Sec:parrot}
In this section, we study the feasibility of creating parrot speech for parrot training. As the parrot speech is the one-shot speech synthesized by a \textcolor{blue}{VC} method, we first introduce the state-of-the-art of VC, then propose a two-step method to generate parrot speech, and finally evaluate how a PT model can approximate a GT model.

\subsection{One-shot Voice Conversion}
\noindent\textbf{Data synthesis:} Generating data with certain properties is commonly used in the image domain, including transforming the existing data via data augmentation \cite{perez2017effectiveness,shorten2019survey,mikolajczyk2018data,shorten2019survey}, generating similar training data via Generative Adversarial Networks (GAN) \cite{goodfellow2020generative,creswell2018generative,aggarwal2021generative}, and generating new variations of the existing data by Variational Autoencoders (VAE) \cite{kingma2013auto,girin2020dynamical,harvey2021conditional,cai2019multi}. These approaches can also be found in the audio domain, such as speech augmentation \cite{ko2015audio,park2019specaugment,ko2017study,li2018training}, GAN-based speech synthesis \cite{chen2020improving,kong2020hifi,kaneko2017generative,binkowski2019high}, and VAE-based speech synthesis \cite{kingma2013auto,hsu2018hierarchical,zhang2019learning}. Specifically, VC \cite{qian2019autovc,liu2021any,lifreevc,wang2021vqmivc,chen2021again} is a specific data synthesis approach that can utilize a source speaker's speech to generate more voice samples that sound like a target speaker. Recent studies \cite{wenger2021hello,deng2023catch} have revealed that it can be difficult for humans to distinguish whether the speech generated by a VC method is real or fake.

\noindent\textbf{One-shot voice conversion:} Recent VC has been developed by only using one-shot speech \cite{chou2019one,lu2019one,wu2020one,chen2021again} (i.e., the methods only knowing one sentence spoken by the target speaker) to convert the source speaker's voice to the target speaker's. This limited knowledge assumption well fits the black-box scenario considered in this paper and motivates us to use one-shot speech data to train a local model for the black-box attacker. As shown in the left-hand side of Fig.~\ref{Fig:parrot_training}, a VC model takes the source speaker's and the target speaker's speech samples as two inputs and yields a parrot speech sample as the output. The attacker can pair the only speech sample, obtained from the target speaker, with different speech samples from public speech datasets as different pairs of inputs to the VC model to generate different parrot speech samples, which are expected to sound like the target speaker's voice to build parrot training.

\subsection{Parrot Speech Sample Generation and Performance}\label{Sec:parrot_speech}
\begin{figure}[t]
    \centering
    \includegraphics[width=0.48\textwidth]{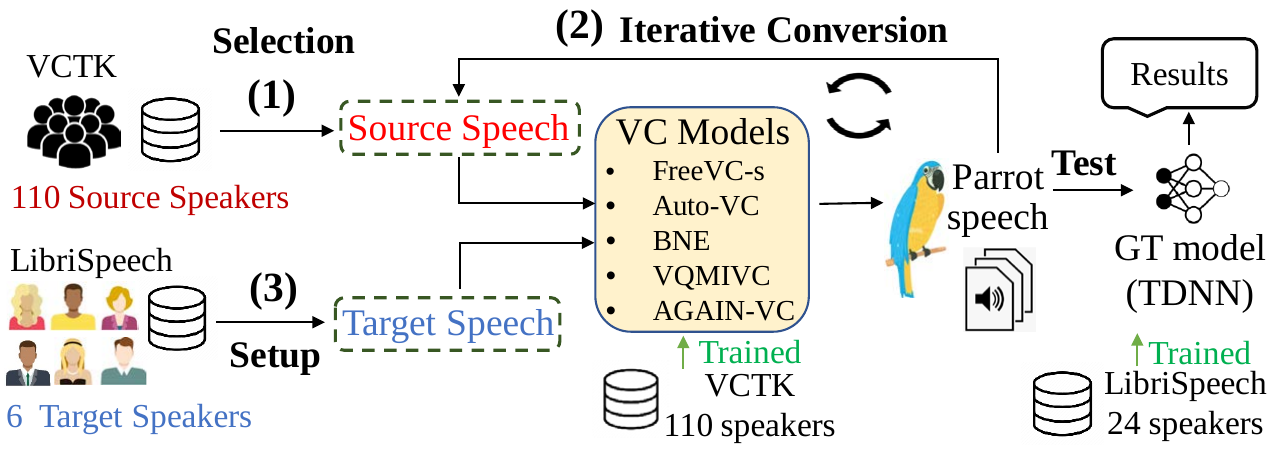}
    \caption{\hlb{Parrot speech generation: setups and evaluations.}}
    \label{Fig:parrot_selection}
    \vspace{-0.4cm}
\end{figure}

We first propose our method to generate parrot speech samples and then use them to build and evaluate a PT model. To generate parrot speech, we propose two design components, motivated by existing results based on one-shot VC methods \cite{kaneko2019cyclegan,kaneko2019stargan, deng2023catch}.
\begin{enumerate}
    \item Initial selection of the source speaker. Existing VC studies \cite{kaneko2019cyclegan,kaneko2019stargan} have shown that intra-gender VC (e.g., female to female) appears to have better performance than inter-gender one (e.g., female to male). As a major difference between male and female voices is the pitch feature \cite{lifreevc,wang2021vqmivc,liu2021any}, which represents the basic frequency information of an audio signal, our intuition is that selecting a source speaker whose voice has the pitch feature similar to the target speaker may improve the VC performance. Therefore, for an attacker that knows a short speech sample of the target speaker to generate more parrot speech samples, the first step in our design is to find the best source speaker in a speech dataset (which can be a public dataset or the attacker's own dataset) such that the source speaker has the minimum average pitch distance to the target speaker.
    \item Iterative conversions. After selecting the initial source speaker, we can adopt an existing one-shot VC method to output a speech sample given a pair of the initial source speaker's and target speaker's samples. As the output sample, under the VC mechanism, is expected to feature the target speaker's audio characteristics better than the initial source speaker, we use this output as the input of a new source speaker's sample and run the VC method again to get the second output sample. We run this process iteratively to eventually get a parrot speech sample. Iterative VC conversions have been investigated in a recent audio forensic study \cite{deng2023catch}, which found that changing the target speakers during iterative conversions can help the source speaker hide his/her voiceprints, i.e., obtaining more features from other speakers to make the voice features of the original source speaker less evident. Compared with this feature-hiding method, our iterative conversions can be considered as a way of amplifying the audio features of the same target speaker to generate parrot speech.
\end{enumerate}
We set up source speaker selection and iterative conversions with one-shot VC models to generate and evaluate the performance of parrot speech samples in Fig.~\ref{Fig:parrot_selection}.

\noindent\textbf{Experimental setup:} There are a wide range of one-shot VC methods recently available for parrot speech generation. We consider and compare the performance of AutoVC \cite{AutoVC}, BNE \cite{PPGVC}, VQMIVC \cite{VQMIVC}, FreeVC-s \cite{FreeVC}, and AGAIN-VC \cite{AGAINVC}. As shown in Fig.~\ref{Fig:parrot_selection}, we use the VCTK dataset \cite{veaux2016superseded} to train each VC model. The dataset includes 109 English speakers with around 20 minutes of speech. We also select the source speakers from this dataset. We select 6 target speakers from the LibriSpeech dataset \cite{panayotov2015librispeech}, which is different from the VCTK dataset, such that the VC training does not have any prior knowledge of the target speaker. Only one short sample (around 4 seconds with 10 English words) of a target speaker is supplied to each VC model to generate different parrot speech samples. We build a time delay neural network (TDNN) as the GT model for a CSI task to evaluate how parrot samples can be accurately classified as the target speaker's voice. The GT model is trained with 24 (12 male and 12 female) speakers from LibriSpeech (including the 6 target speakers and 18 randomly selected speakers). The model trains 120 speech samples (4 to 15 seconds) for each speaker and yields a test accuracy of 99.3\%.

\noindent\textbf{Evaluation metrics:} We use the False Positive Rate (FPR) \cite{huang2021stop,chen2019real} to evaluate the effectiveness of parrot speech, i.e., the percentage of parrot speech samples that are classified by the TDNN classifier as the target speaker's voice. Specifically, $\text{FPR}=  \text{FP} / (\text{FP} + \text{TN}) $, where False Positives (FP) indicates the number of cases that the classifier wrongly identifies parrot speech samples as target speaker's label; True
Negatives (TN) represents the number of cases that the classifier correctly rejects parrot speech samples as any other label except for the target speaker.

\begin{figure}[t!]
   \begin{minipage}[t]{0.495\columnwidth}
     \centering
     \includegraphics[width=\linewidth]{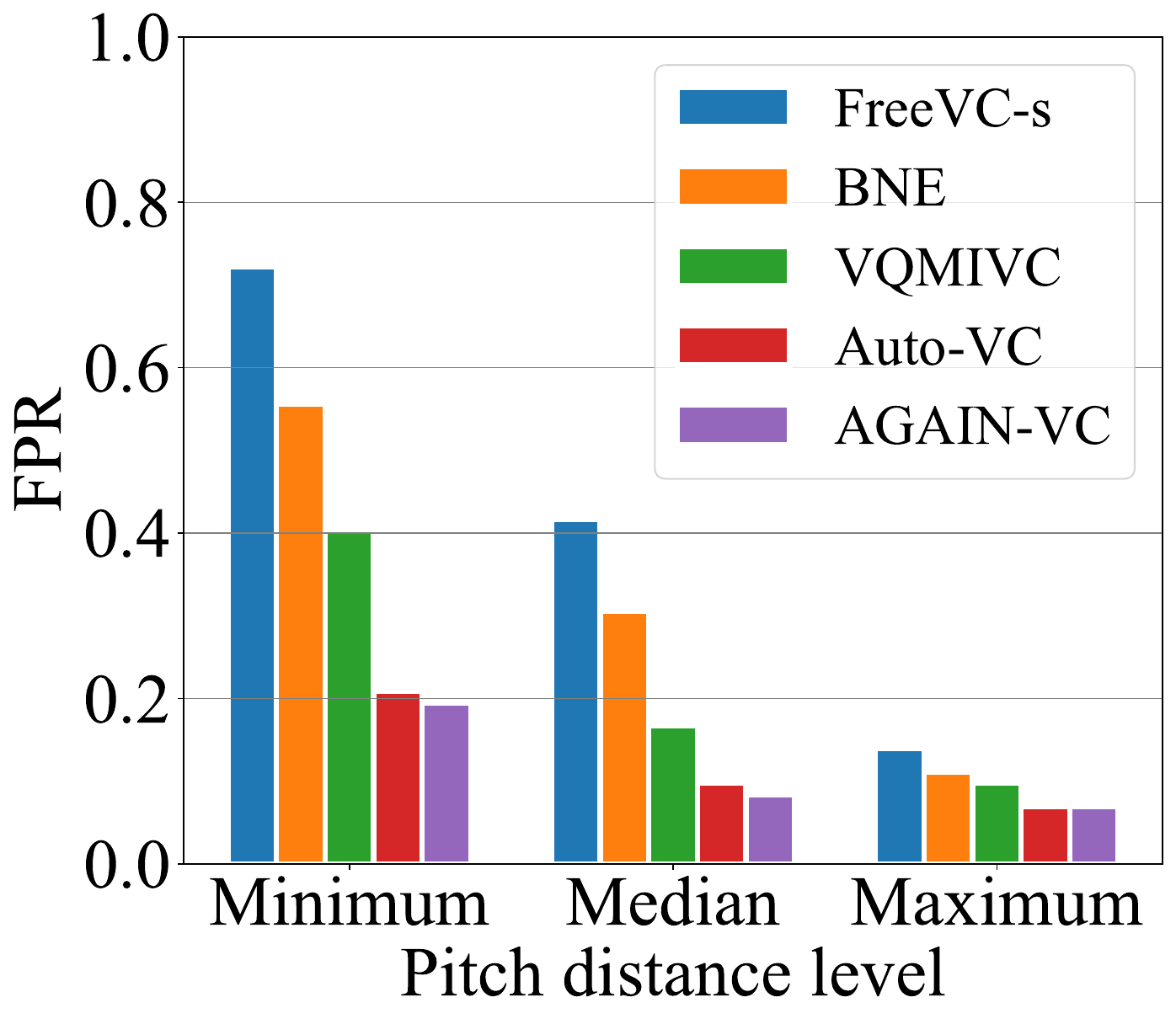}
     \caption{\hlb{FPRs under different initial source speakers.}}\label{Fig:pitch_distance}
   \end{minipage}
   \begin{minipage}[t]{0.495\columnwidth}
     \centering
     \includegraphics[width=\linewidth]{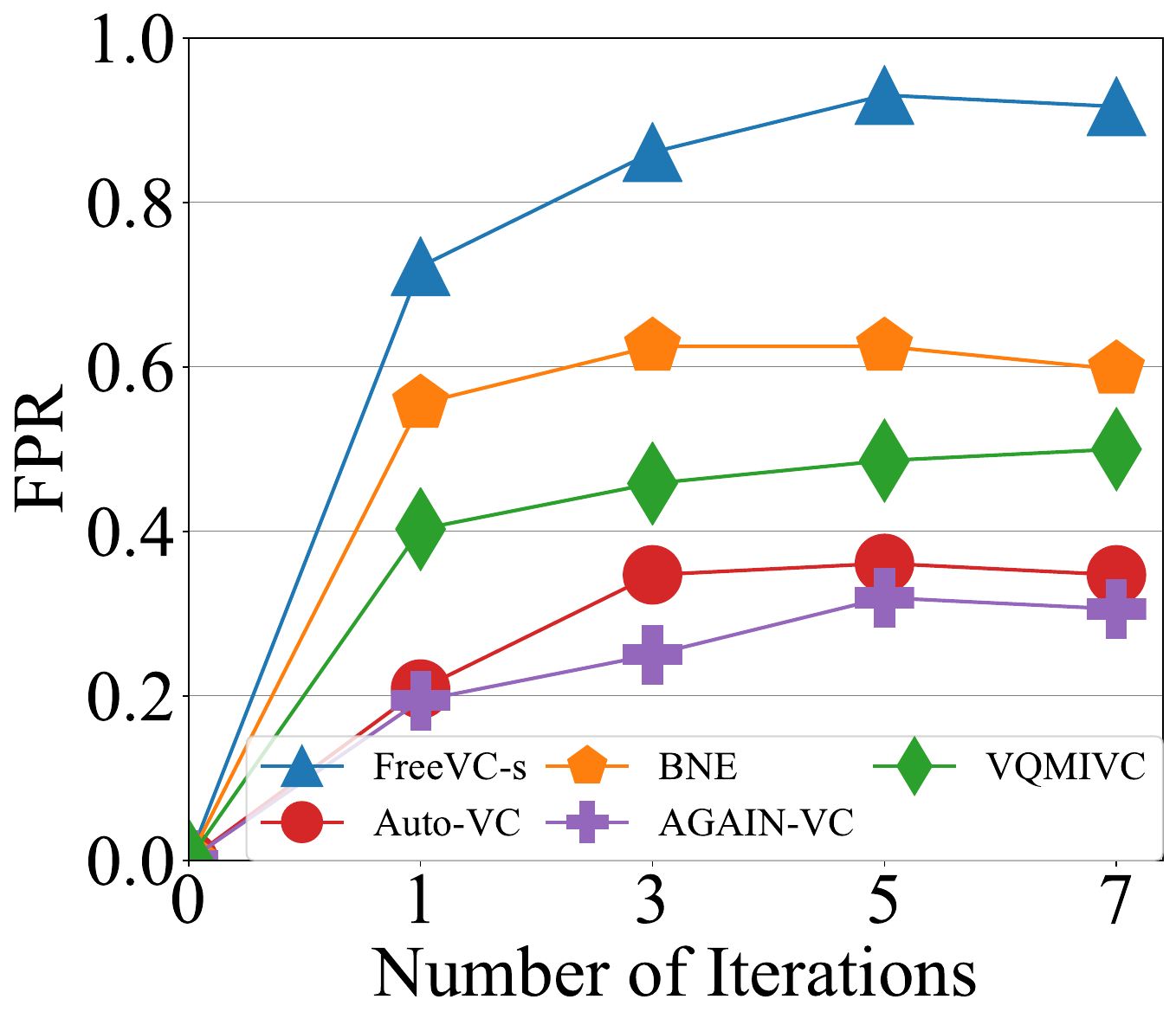}
     \caption{\hlb{FPRs under different numbers of iterations.}}\label{Fig:iterative}
   \end{minipage}
   \vspace{-0.4cm}
\end{figure}

\noindent\textbf{Evaluation results:} We first evaluate the impact of the initial source speaker selection on different VC models. We set the number of iterative conversions to be one, and the target speaker's speech sample is around 4.0 seconds (10 English words), which is the same for all VC models. We use the pitch distance between the source and target speakers as the evaluation standard. Specifically, we first sort all 110 source speakers in the VCTK dataset with respect to their average pitch distances to the target speaker. We use \textit{minimum}, \textit{median}, \textit{maximum} to denote the source speakers who have the smallest, median, and largest pitch distances out of all the source speakers, respectively.  We use each VC method to generate 12 different parrot speech samples for each target speaker (i.e., a total of 72 samples for 6 target speakers under each VC method). Fig.~\ref{Fig:pitch_distance} shows that the pitch distance of the source speaker can substantially affect the FPR. For the most effective VC model, Free-VCs, we can observe that the FPR can reach 0.7222 when the source speaker is chosen to have the minimum distance to the target speaker, indicating that 72.22\% parrot speech samples can fool the GT TDNN model in Fig.~\ref{Fig:parrot_selection}. Even for the worst-performing {\hlb AGAIN}-VC model, we can still observe that the minimum-distance FPR (0.1944) is nearly 3 times the maximum-distance FPR (0.0694). As a result, the source speaker with the less pitch distance is more effective to improve the VC performance (i.e., leading to a higher FPR).

Next, we evaluate the impact of iterative conversions on the FPR. Fig.~\ref{Fig:iterative} shows the FPRs with different numbers of iterations for each VC model (with zero iteration meaning no conversion and directly using the TDNN to classify each source speaker's speech). It is noted from the figure that with increasing the number of iterations, the FPR initially gains and then stays within a relatively stable range. For example, the FPR of FreeVC-s achieves the highest value of 0.9305 after 5 iterations and then drops slightly to 0.9167 after 7 iterations. Based on the results in Fig.~\ref{Fig:iterative}, we set 5 iterations for parrot speech generation.

We are also interested in how much knowledge of the target speaker is needed for each VC model to generate effective parrot speech. We set the knowledge level based on the length of the target speaker's speech given to the VC. Specifically, we crop the target speaker's speech into four levels: i) 2-second length level (around 5 words), ii) 4-second level (10 words), iii) 8-second level (15 words), and iv) 12-second level: (22 words). For each VC model, we generate 288 parrot speech samples (12 for each target speaker with each different knowledge level) to interact with the GT model. All samples are generated by choosing the initial source speaker with the minimum pitch distance and setting the number of iterations to be 5.

\begin{table}[t]
\centering
\caption{VC Performance under different knowledge levels.}
\begin{adjustbox}{center,max width=1\linewidth}
\begin{tabular}{lccccc}
\toprule
\begin{tabular}[l]{@{}l@{}}Knowledge \\ Level\end{tabular} & FreeVC-s & AutoVC & BNE     & VQMIVC & {\hlb AGAIN}-VC \\\hline 
2-second                                                     & 0.5416 & 0.0972 & 0.3194  & 0.1667 & 0.0833   \\ 
4-second                                                    & 0.8750 & 0.4028 & 0.5139  & 0.4583 & 0.2639   \\ 
8-second                                                       & 0.9167 & 0.5417 & 0.7083 & 0.5833 & 0.3750   \\ 
12-second                                                       & 0.9305 & 0.5556 & 0.7222 & 0.5972 & 0.3889   \\ \bottomrule
\end{tabular}
\label{Tab:VC_knowledge}
\end{adjustbox}
\vspace{-0.4cm}
\end{table}

Table~\ref{Tab:VC_knowledge} evaluates the FPRs under different knowledge levels of the target speaker. It can be seen that the length of the target speaker's speech substantially affects the effectiveness of parrot speech samples. For example, AutoVC achieves the FPRs of 0.0972 and 0.5417 given 2- and 4-second speech samples of the target speaker, and finally increases to 0.5556 with the 12-second knowledge. It is also observed that FreeVC-s performs the best in all VC methods for each knowledge level (e.g., 0.9167 for the 8-second knowledge level). We can also find that the increase in FPR becomes slight from 8-second to 12-second speech knowledge. For example, FreeVC-s increases from 0.9167 (8-second) to 0.9305 (12-second), and VQMIVC increases from 0.5833 (8-second) to 0.5972 (12-second).
Overall, the results of Table~\ref{Tab:VC_knowledge} reveal that even based on a very limited amount (i.e., a few seconds) of the target speaker's speech, parrot speech samples can still be efficiently generated to mimic the speaker's voice features and fool a speaker classifier to a great extent.

\subsection{Parrot Training Compared with Ground-Truth Training}
We have shown that parrot speech samples can be effective in misleading a GT-trained speaker classification model. Additionally, we use experiments to further evaluate how a PT model trained by parrot speech samples is compared with a GT model. 
{\hlb We compare the classification performance of PT and GT models. Based on our findings, PT models exhibit classification performance that is comparable to, and can approximate, GT models. We include experimental setups and results in Appendix~\ref{sec:PTvsGT}.}

\section{PT-AE Generation: A Joint Transferability and Perception Perspective}\label{Sec:perception}
In this section, we aim to evaluate whether the PT-AEs are as effective as GT-AEs against a black-box GT model. We first summarize AE generation methods that use different types of audio waveforms (i.e., carriers). Next, we quantify the human perceptual quality of AEs with different carriers, then use the match rate to measure the transferability of PT-AEs to GT models. Finally, we define the unified metric, transferability-perception ratio (TPR), to evaluate PT-AEs.

\subsection{Carriers in Audio AE Generation}
Recent audio attack studies have considered different audio perturbation carriers to generate AEs via specific generation algorithms. We summarize three main types of carriers.

\noindent{\bf{Noise carriers:}} Traditional methods \cite{chen2019real,liu2022evil} usually adopt a gradient estimation method to generate audio AEs in the unrestricted $L_p$ space with the initial perturbation signal set commonly as a Gaussian noise. This leads to a noisy sound despite some psychoacoustic methods \cite{qin2019imperceptible,guo2022specpatch,liu2022evil} that can be used to alleviate the noisy effect.

\noindent{\bf{Feature-twisted carriers:}} Directly manipulating the auditory feature of a speech signal could make a classifier sensitive but stealthy to the human ears. Existing works \cite{abdullah2019hear,yusmack} have found that modifying the phonemes or changing the prosody of the speech can also spoof the audio classifier while preserving the perception quality.  

\noindent{\bf{Environmental sound carriers:}} The enrollment phase attack \cite{deng2022fencesitter} employed environmental sounds (e.g., traffic) to create the perturbation signal to poison a speaker recognition model. 

\subsection{Quantifying Perceptual Quality of Speech AEs}\label{Sec:Human_study}
We first need to find an appropriate perception metric to accurately measure the human perceptual quality of AEs based on different carriers. Recent studies \cite{duan2022perception,yusmack} have pointed out that traditional metrics, such as signal-to-noise ratio (SNR) \cite{chen2020devil} and the $L_p$ norm \cite{yuan2018commandersong,chen2019real,zheng2021black}, cannot directly reflect the human perception. They have used different human study based metrics to measure the perceptual quality of AEs with certain types of carriers (i.e., qDev for music AEs in \cite{duan2022perception} and NISQA for feature-twisted AEs \cite{yusmack}). In addition, we also notice that the harmonics-to-noise ratio (HNR) \cite{yumoto1982harmonics} is a common metric adopted in speech science to measure the quality of a speech signal. Given these potential perception metrics, we aim at conducting a human study to find out the best metric to measure the perceptual quality across a diversity of AE carriers that we are interested in.

\noindent\textbf{Dataset generation for human study:}
We create the human study dataset with noise carriers \cite{carlini2018audio, qin2019imperceptible, guo2022specpatch, chen2019real, zheng2021black, liu2022evil}, feature-twisted carriers \cite{yusmack}, and environmental carriers \cite{deng2022fencesitter}. We choose 30 original speech signals (with length from 5 to 15 seconds) from the existing speech dataset \cite{mittag2021nisqa}. We modify these original signals by adding different types of carriers to form perturbed speech signals for the human study. We use the signal-to-carrier ratio (SCR) to control the energy of a perturbation carrier added to an original signal. For example, an SCR of 0dB means that the carrier and the original signal have the same energy level.
We consider the following carriers to be added to the original signals.

\noindent{i) Noise carriers:} The dataset \cite{mittag2021nisqa} provides a wide range of noisy speech signals. The noise is Gaussian-distributed and can be generated with different SNRs. We generated 30 speech samples whose SNRs are uniformly distributed in 0-30 dB. Note that the metric SCR is equivalent to the metric of SNR in the case of noise carriers.

\noindent{ii) Feature-twisted carriers:} For feature-twisted speech signals, we shift the tone (i.e., the pitch) \cite{yusmack} to generate pitch-twisted carriers. Specifically, we shift up/down by 25 semitones\footnote{1 semitone = $12\log_2(f^\prime / f)$, where $f$ and $f^\prime$  are the original and perturbed speech frequencies, respectively \cite{semitone}.} of the original speech to craft the pitch-twisted carriers, and add these carriers to the original speech with different SCR levels. For twisting the rhythm, we speed up and slow down the speech ranging from 0.5 to 2 times of its speech rate. 

\noindent{iii) Environmental sound carriers:} Environmental sound carriers are selected from the large-scale human-labeled environmental sound datasets \cite{gemmeke2017audio} with categories including natural sounds (e.g., wind and sea waves), sounds of things (e.g., vehicle and engine), human sounds (e.g., whistling), animal sounds (e.g., pets), and music (e.g., musical instruments). For each category, we randomly selected 6 audio clips.

We have created a total of 90 perturbed speech samples, 30 samples for each carrier set at different SCR levels.

\noindent\textbf{Human participant involvement:} We have recruited 30 volunteers, who are college students with no hearing issues (self-reported). Our study procedure was approved by our Institutional Review Board (IRB). Each volunteer is asked to rate the similarity between a pair of original and carrier-perturbed speech clips using a scale from 1 to 7 commonly adopted in speech evaluation studies \cite{gemmeke2017audio,11wester,bunton2007listener,anand2019objective,patel2010perceptual,darley1969differential}, where 1 indicates the least similarity (i.e., speakers sound very different between the two clips) and 7 represents the most similarity (i.e., speakers sound very similar).

\begin{figure}[t]
    \centering
    \includegraphics[width=0.42\textwidth]{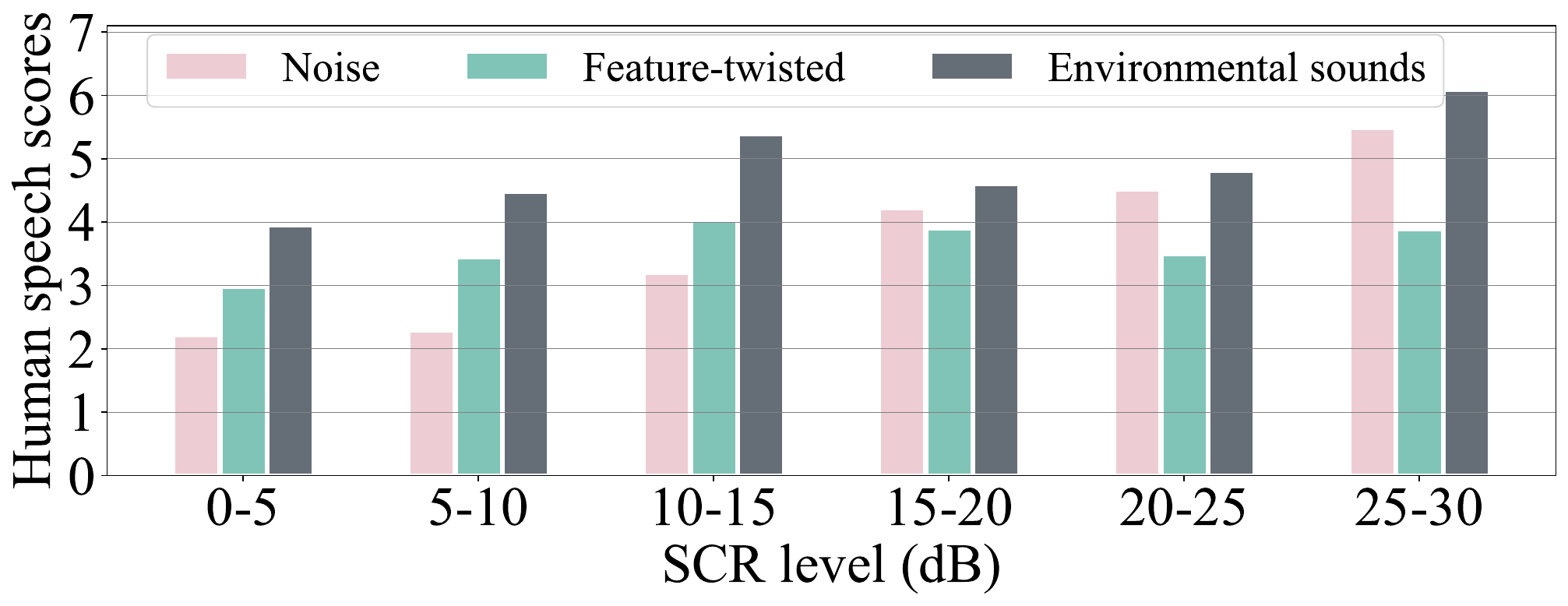}
    \caption{Human scores for carrier-perturbed speech signals.}
    \vspace{-0.0cm}
    \label{Fig:human_ratings_noise_vs_carrier}
    \vspace{-0.6cm}
\end{figure}

\noindent\textbf{Perceptual quality of different carriers:} Fig.~\ref{Fig:human_ratings_noise_vs_carrier} compares the average human scores at varying SCR levels for different carriers. We can clearly see that the perception quality for noise carriers improves gradually with increasing the SCR, which indicates the less loudness of the noise carrier, the better perception of the perturbed speech. Interestingly, the human scores of the feature-twisted and environmental sound carriers are not closely correlated with the SCR. Both of them can indeed get better human scores at lower SCR levels (e.g., 10-15 dB vs 15-20 dB). Fig.~\ref{Fig:human_ratings_noise_vs_carrier} also shows that overall, environmental sound carriers yield the better human scores than the feature-twisted carriers and noise carriers.

\begin{table}[t]
\centering
\caption{Evaluation of different metrics.}
\label{tab:Pearson_and_Spearman_deviation measure}
\begin{adjustbox}{center,max width=0.88\linewidth}
\begin{tabular}{cccccccc}
\hline
Carrier Type                                                                   & Metrics  & SRS   & HNR    & $L_2$      & $L_\infty$      & SCR    & NISQA  \\ \hline
\multirow{2}{*}{Noise}                                                         & Pearson  & \textbf{0.9387} & 0.6339 & -0.7699 & -0.6680 & 0.2524 & 0.9279 \\
                                                                               & Spearman & 0.7882 & 0.7303 & \textbf{-0.9349} & -0.9229 & 0.3956 & 0.8409 \\ \hline
\multirow{2}{*}{\begin{tabular}[c]{@{}c@{}}Environ. \\ Sounds\end{tabular}} & Pearson  & \textbf{0.9647} & 0.4265 & 0.0923  & -0.5426 & 0.2348 & 0.6657 \\
                                                                               & Spearman & \textbf{0.9566} & 0.5355 & -0.2843 & -0.4761 & 0.4152 & 0.7280 \\ \hline
\multirow{2}{*}{\begin{tabular}[c]{@{}c@{}}Feature-\\ twisted\end{tabular}}                                                      & Pearson  & \textbf{0.9234} & 0.1099 & -0.1959 & 0.0744  & -0.097 & 0.3859 \\
                                                                               & Spearman & \textbf{0.9139} & 0.1173 & -0.0985 & -0.0097 & 0.0397 & 0.2978 \\ \hline
\multirow{2}{*}{Overall}                                                       & Pearson  & \textbf{0.9299} & 0.0855 & -0.3108 & -0.4068 & 0.0438 & 0.2372 \\
                                                                               & Spearman & \textbf{0.9187} & 0.0785 & -0.3691 & -0.4603 & 0.1331 & 0.1434 \\ \hline
\end{tabular}
\end{adjustbox}
\vspace{-0.3cm}
\end{table}

\noindent\textbf{Evaluation of speech quality metrics:} Next, we evaluate the accuracy of existing metrics to characterize the speech quality based on our human study results. We compare the metrics of $L_2$ and $L_\infty$ norms \cite{yuan2018commandersong,chen2019real,zheng2021black}, SCR (equivalent to SNR \cite{chen2020devil}), HNR \cite{yumoto1982harmonics}, audio-feature-regression-based qDev \cite{duan2022perception}, and DNN-based NISQA \cite{yusmack,mittag2021nisqa}. Note that the qDev model \cite{duan2022perception} was originally trained using music instead of speech. We follow the procedure in \cite{duan2022perception} to train a random forest regression model using our speech samples. We call the resultant metric speech-regression score (SRS).

To evaluate how well a speech quality metric matches the human score from the human study, we use two correlation coefficients, Pearson's and Spearman's coefficients \cite{hauke2011comparison}, to measure the correlation between the metric and the human score. Table~\ref{tab:Pearson_and_Spearman_deviation measure} computes all correlation coefficients from our human study. It is observed from the table that SRS has the best accuracy across almost all carriers, except for noise carriers, where $L_2$-norm achieves the highest Spearman's coefficient. The DNN-based NISQA has high coefficients for noise carriers, but has degraded accuracy for feature-twisted carriers. One potential reason is that NISQA is trained with the noise carrier and environmental sound carrier dataset \cite{mittag2021nisqa}, which may not be effective for feature-twisted speech as the diversity of training data is important to the prediction performance \cite{duan2022perception}. Based on Table~\ref{tab:Pearson_and_Spearman_deviation measure}, we use the metric of SRS to measure the perpetual quality of an audio AE.

\subsection{Measuring Transferability of PT-AEs}\label{Sec:Measuring_PT-AE}
We then move to evaluate the transferability of different carriers for PT-based AEs.

\subsubsection{Building Target and Surrogate Models}
The first step in evaluating the transferability is to build i) target models, which refer to the models to be attacked by the attacker using PT-AEs, and ii) surrogate models, which are used by the attacker to generate PT-AEs against the target models. It is known that the difference between the target and surrogate models can affect the transferability of AEs \cite{liu2016delving}.

\noindent\textbf{Building target models:} We consider building a diversity of target models with 4 DNN-based speaker recognition models including 2 CNN \cite{jati2021adversarial} and 2 TDNN models \cite{snyder2017deep,snyder2018x}. These 4 target models are trained with the same 6 target speakers (3 males and 3 females). We randomly select them from LibriSpeech, and use 120 speech samples for each speaker for training. As the 4 target models have varying architectures and parameters (i.e., number of layers and weights), we denote them as CNN-A, CNN-B, TDNN-A, and TDNN-B. Their accuracies are 100.0\%, 96.5\%, 99.3\%, and 97.2\%, respectively.

\noindent\textbf{Building surrogate models:} We also aim to build a diversity of
surrogate models for the attacker. As the attacker, without the knowledge of target models, is free to use any architecture for parrot training, we build two CNN-based and two TDNN-based surrogate architectures with different parameters, denoted by PT-CNN-C, PT-CNN-D, PT-TDNN-C, and PT-TDNN-D. Since there are 6 speakers trained in a target model, we consider each of them to be the attacker's target under each of the four surrogate architectures. For example, when the attacker uses the PT-CNN-C architecture and she targets speaker $i\in [1,6]$ in the target models, the attacker is assumed to only know speaker~$i$'s 8-second speech, and uses it to generate parrot speech samples, together with speech samples from 3 to 8 speakers randomly selected from the VCTK dataset (none is in the target models that use the LibriSpeech dataset), to build her surrogate model, denoted by PT-CNN-C-$i$. As a result, we construct a set of 6 surrogate models under each surrogate architecture (totally 24 models), denoted by \{PT-CNN-C-$i$\}$_{i\in[1,6]}$, \{PT-CNN-D-$i$\}$_{i\in[1,6]}$, \{PT-TDNN-C-$i$\}$_{i\in[1,6]}$, and \{PT-TDNN-D-$i$\}$_{i\in[1,6]}$.

\noindent\textbf{Compare PT with benchmark GT models.} To better understand the transferability of the PT-AEs in comparison with GT-AEs, we also use the target speaker~$i$'s ground-truth speech instead of the parrot speech to build the attacker's surrogate models under the four surrogate architectures, denoted by \{GT-CNN-C-$i$\}$_{i\in[1,6]}$, \{GT-CNN-D-$i$\}$_{i\in[1,6]}$, \{GT-TDNN-C-$i$\}$_{i\in[1,6]}$, and \{GT-TDNN-D-$i$\}$_{i\in[1,6]}$. We will also generate GT-AEs based on these GT-surrogate models to attack the target models. They will serve as the benchmark for comparison with their PT counterparts.

\subsubsection{AE generations via different carriers}\label{Sec:AE_generation} After building the surrogate and target models, we generate AEs from the surrogate models using the three types of carriers based on existing studies.

\vspace{-0.1cm}
\noindent i) For the noise carrier, we solve the white-box problem \eqref{Eq:BasicWhiteBox} via projected gradient descent (PGD) \cite{goldstein2014field}, and we choose $L_{\infty}$ norm as the distance metric, which shows a good performance in Table~\ref{tab:Pearson_and_Spearman_deviation measure}. We set $\epsilon=0.05$ to control the $L_{\infty}$ norm.

\vspace{-0.1cm}
\noindent ii) For the feature-twisted carrier, we twist the pitch and rhythm of the original speech \cite{yusmack,duan2022perception} using the perception metric SRS as the distance measurement. As the random-forest-based SRS is non-differentiable, we use grid search to solve \ref{Eq:BasicWhiteBox}. Specifically, we shift up/down for 25 semitones of the pitch, and the minimal shift-pitch step $\Delta_p=1$ semitone. We speed up and slow down the speech ranging from 0.2 to 2.0 its speech rate with the minimal rhythm-changed step $\Delta_r$ to be 0.2.

\vspace{-0.1cm}
\noindent iii) For the environmental sound carrier, we choose 30 environmental sounds from \cite{gemmeke2017audio} which includes natural sounds, sounds of things, human sounds, animal sounds, and music. Based on the SRS to represent the distance $D$ in \eqref{Eq:BasicWhiteBox}, we solve \eqref{Eq:BasicWhiteBox} via finding the best linear weights \cite{duan2022perception} of different environmental sounds using grid search with the minimal search step to be 0.1$\epsilon$ with threshold $\epsilon$ set to be 0.05 (the same as the noise carrier's threshold).

For each carrier type, we generate 20 PT-AEs from each PT-surrogate model (a total of 480 PT-AEs). In addition, we generate 20 GT-AEs from each GT-surrogate model for the comparison purpose (also a total of 480 GT-AEs).

\begin{table*}[t]
    \centering
    \caption{Match rates between surrogate and target models.}
    \begin{adjustbox}{center,max width=0.86\linewidth}
    \begin{tabular}{r|ccccc|ccccc|ccccc}
    \toprule
    AE Carrier Type:  & \multicolumn{5}{c|}{Noise}                            & \multicolumn{5}{c|}{Feature-twisted}                  & \multicolumn{5}{c}{Environmental sound}              \\\midrule
     Target Model:   & CNN-A   & CNN-B   & TDNN-A  & TDNN-B  & \textbf{Average}& CNN-A   & CNN-B   & TDNN-A  & TDNN-B  & \textbf{Average} & CNN-A   & CNN-B   & TDNN-A  & TDNN-B  & \textbf{Average}
    
    \\ \midrule
    GT-CNN-C     & 0.2167      & 0.1500 & 0.1167 & 0.1417 & \textbf{0.1563}  & 0.2333 & 0.2083 & 0.1583 & 0.1750 & \textbf{0.1937}  & 0.3500 & 0.3250 & 0.2417      & 0.2250  & \textbf{0.2854}
    
    \\ \midrule
    PT-CNN-C     & 0.1917      & 0.1417 & 0.0917 & 0.1250 & \textbf{0.1375}  & 0.2083  & 0.1750 & 0.1083 & 0.1583 & \textbf{0.1625}   & 0.3083 & 0.2583 &0.2000      & 0.1750 & \textbf{0.2353}
    \\ \midrule
    
    GT-CNN-D    & 0.0917 & 0.2167      & 0.0833 & 0.1917 & \textbf{0.1458}  & 0.1667 & 0.1917     & 0.1500 & 0.1833 & \textbf{0.1729}  & 0.1833      & 0.3250 & 0.2417 & 0.2917 & \textbf{0.2604} \\ \midrule
    
    PT-CNN-D    & 0.0417 & 0.1667      & 0.0583 & 0.1583 & \textbf{0.1063}  & 0.1417 &0.1500  & 0.1417 & 0.1583 & \textbf{0.1479}  & 0.1583       & 0.2167& 0.2750 & 0.2583 & \textbf{0.2271}
    \\ \midrule
    
    GT-TDNN-C   & 0.1000 & 0.1500 & 0.1750      & 0.1583 & \textbf{0.1458}  & 0.1500 & 0.1833 & 0.2583     & 0.1417 & \textbf{0.1833}  & 0.3500 & 0.1833 & 0.3583      & 0.3417 & \textbf{0.3083} \\ \midrule
    
    PT-TDNN-C   & 0.0917 & 0.1417 & 0.1667      & 0.1333 & \textbf{0.1333}  & 0.1167 & 0.1750 & 0.2500     & 0.1333 & \textbf{0.1688}  & 0.3167 & 0.1750 & 0.2833      & 0.3083 & \textbf{0.2708}
    \\ \midrule
    
    GT-TDNN-D    & 0.1333 & 0.1000 & 0.2083 & 0.2083     & \textbf{0.1625}  & 0.1583 & 0.2750 & 0.2833 & 0.2917      & \textbf{0.2520}  & 0.1417 & 0.3083 & 0.3917 & 0.4083      & \textbf{0.3125} \\ \midrule
    
    PT-TDNN-D    & 0.1250 & 0.0833 & 0.1750 & 0.1667      & \textbf{0.1375}  & 0.1417 & 0.2500 & 0.2500 & 0.2583      & \textbf{0.2225}  & 0.1250 & 0.2667 & 0.3417 & 0.3333      & \textbf{0.2667}  \\ \hline
    \end{tabular}
    \end{adjustbox}
    \label{Tab:3method_transfer}
    \end{table*}

\subsubsection{Evaluation metric for transferability}
The transferability has been extensively studied in the image domain \cite{papernot2016transferability,liu2016delving,papernot2017practical,mao2022transfer}. One important evaluation metric in the transfer attacks \cite{mao2022transfer,liu2016delving} is the match rate, which measures the percentage of AEs that can make both a surrogate model and a target model predict the same wrong label. We use the metric of the match rate to measure the transferability of PT-AEs in this work. Specifically, we can test a generated PT-AE: $x+\delta$ with both surrogate model $f(\cdot)$ and target model $f^\prime(\cdot)$. If $f(\!x+\!\delta) \!=\! f^\prime(x\!+\!\delta) \neq f(x) $, we can say $x\!+\!\delta$ is a matched AE for both $f(\cdot)$ and $f^\prime(\cdot)$. The match rate is the ratio between the number of matched AEs and the total number of AEs.

\subsubsection{Results analysis}
It would be tedious to show the match rate of each pair in the 24 surrogate models and 6 target models that we have built. We average the match rates of the surrogate models under the same surrogate architecture (i.e., PT-CNN-C, PT-CNN-D, PT-TDNN-C, and PT-TDNN-D). For example, we compute the match rate of the PT-CNN-C based surrogate architecture by averaging the six match rates of \{PT-CNN-C-$i$\}$_{i\in[1,6]}$ models against a target model.

Table~\ref{Tab:3method_transfer} shows the match rates between different surrogate and target models under the 3 types of AE carriers. We can see that the environmental sound carrier achieves better AE transferability than the noise and feature-twisted carriers in terms of the average match rate over the 4 target models. In particular, PT-AEs based on environmental sounds have match rates from 0.23 to 0.27, compared with 0.10 to 0.14 (noise carrier) and 0.15 to 0.22 (feature-twisted carrier). The results demonstrate that using environmental sounds as the carrier achieves the best transferability of PT-AEs from a PT-surrogate model to a target model.

Table~\ref{Tab:3method_transfer} also compares the match rates of PT-AEs generated from PT models in comparison with GT-AEs generated from GT models. We can observe that the match rate of PT-AEs is slightly lower than their GT counterparts. For example, using the noise carrier, GT-AEs based on GT-TDNN-D achieve the best average match rate of 0.1625; in contrast, PT-AEs based on PT-TDNN-D obtain a slightly lower average match rate of 0.1375. Overall, we can see that PT-AEs are slightly less transferable than GT-AEs, but still effective against target models, especially using the environmental sound carrier.


\subsection{Defining Transferability-Perception Ratio for Evaluation}\label{Sec:TPR_metric}
Now, given an AE carrier type $C \in$ \{\text{noise}, \text{feature-twisted}, \text{environmental sounds}\}, we have the metrics of SRS($C$) and match rate $m(C)$ to measure the perceptual quality and transferability of PT-AEs of type $C$, respectively. We define a joint metric, named Transferability-Perception Ratio (TPR), as
\begin{equation}
\text{TPR}(C)= m(C) / (8 - \text{SRS}(C)),
\end{equation}
where $8 - \text{SRS}(C)$ ranges from 1 to 7, denoting the score loss to the best human perceptual quality. The resultant value of $\text{TPR}(C)$ is in $[0,1]$ and quantifies, on average, how much transferability (in terms of the match rate) we can obtain by degrading one unit of human perceptual quality (in terms of the SRS). A higher TPR indicates a better AE quality from a joint perspective of transferability and perception.

As the attacker only knows one-sentence speech of her target speaker, the length of the speech (measured by seconds) is an important factor for the attacker to build the PT model and determines the effectiveness of PT-AEs. Fig.~\ref{Fig:TPR1} shows the TPRs of PT-AEs using the 3 types of carriers under different attack knowledge levels (2, 4, 8, and 12 seconds). It is observed in Fig.~\ref{Fig:TPR1} that the TPRs of all AE carriers increase by giving more knowledge about the target speaker's speech. For example, the TPR of the environmental sound carrier increases substantially from 0.14 (4-second level) to 0.25 (8-second level), and then slightly to 0.259 (12-second level).

Note that the environmental sound carrier in all three types has the highest TPR at each knowledge level, which is consistent with the findings in Fig.~\ref{Fig:human_ratings_noise_vs_carrier} and Table~\ref{Tab:3method_transfer}. We also see that the feature-twisted carrier achieves the second-highest TPR, while the noise carrier has the lowest TPR. In summary, our TPR results show that we can base environment sounds to generate PT-AEs to improve their transferability to a black-box target model.

\section{Optimized Black-box PT-AE Attacks}\label{Sec:ensemble}
In this section, we propose an optimized PT-AE generation mechanism to attack a black-box target model. We first investigate the TPRs of PT-AEs generated from combined carriers, then formulate a two-stage attack to generate PT-AEs against the target model.

\subsection{Combining Carriers for Optimized PT-AEs}\label{sec:combining_carriers}
\begin{figure}[t]
    \centering
    \includegraphics[width=0.45\textwidth]{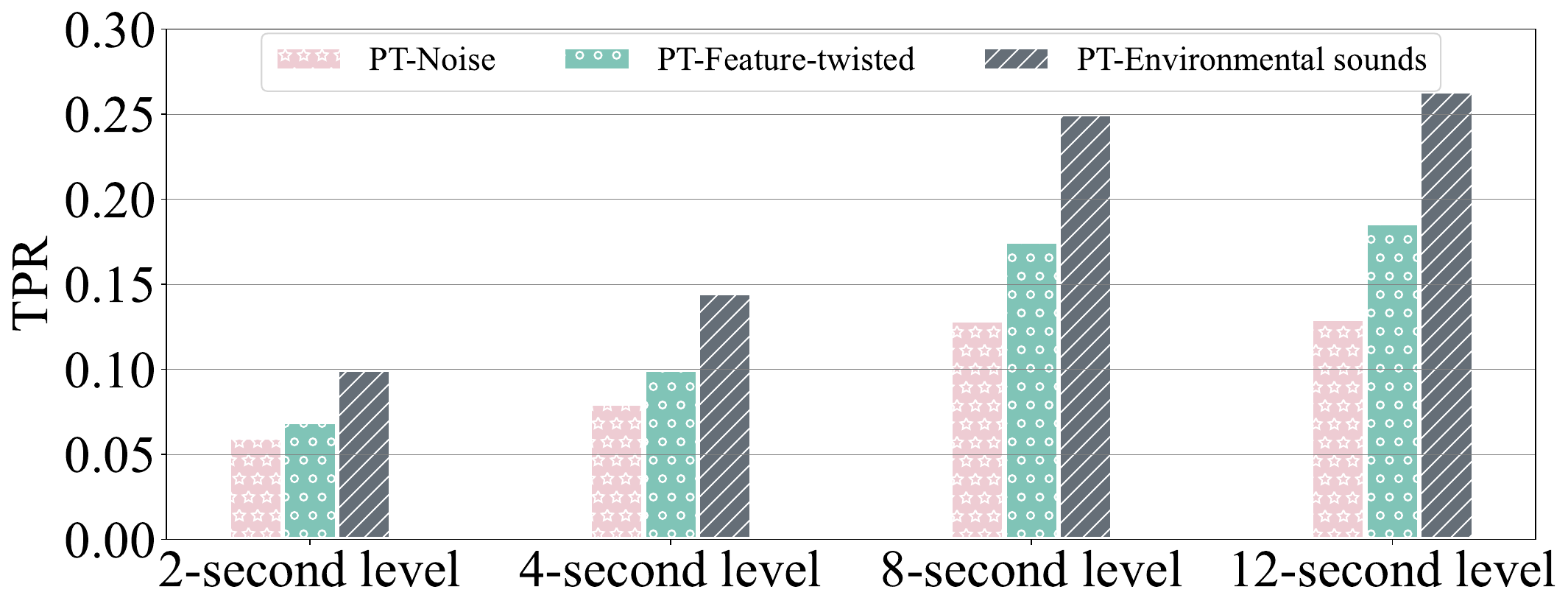}
    \caption{TPRs of carriers with different attack knowledge levels.}
    \label{Fig:TPR1}
    \vspace{-0.5cm}
\end{figure}

The findings in Fig~\ref{Fig:TPR1} reveal that the environmental sound carrier achieves the highest TPR and should be a good choice to generate PT-AEs. But using the environmental sound carrier does not exclude us to further twist the auditory feature of the carrier or adding additional noise to it (e.g., an enrollment-phase attack \cite{deng2022fencesitter} used both environmental sounds and noise). In other words, there is a potential way to combine the environmental sound carrier with feature-twisting or noise-adding method to further improve the TPR.

We consider two additional types of carriers: (i) Feature-twisted environmental sounds, and manipulating the pitch \cite{yusmack} or the rhythm \cite{duan2022perception} is a straightforward way to twist the features of environmental sounds. We follow the same feature-twisting procedure in Section~\ref{Sec:AE_generation} to twist the pitch and rhythm features of environmental sounds to generate PT-AEs. (ii) Noise-based environmental sounds. We first add environmental sounds to the original speech and then use the noise attack procedure in Section~\ref{Sec:AE_generation} to generate PT-AEs.

Fig.~\ref{Fig:twisted_feature_TPR} shows the TPRs of various PT-AEs generated based on (i) adding noise to, (ii) twisting the rhythm, and (iii) twisting the pitch of a type of environmental sounds. We can find that the TPR is sensitive to the choice of environmental sounds. For example, the music sounds do not seem very effective to increase the TPRs even with twisted features. It is noted that natural sounds have overall higher TPRs than other types of carriers. For example, using the brook sounds can achieve 0.29 TPR compared with alarm (0.25), rooster (0.26), and Rock2 (0.16) in the existing dataset \cite{gemmeke2017audio}. Moreover, Fig.~\ref{Fig:twisted_feature_TPR} illustrates the uniform advantage of twisting the pitch of environmental sound over twisting the rhythm and adding noise. For example, built upon the hail sounds, twisting the pitch feature obtains a TPR of 0.26, substantially higher than twisting the rhythm (0.18) and adding noise (0.05). In addition, Fig.~\ref{Fig:twisted_feature_TPR} shows that adding noise is the least effective way to improve the TPR. Based on the results in Fig.~\ref{Fig:twisted_feature_TPR}, we consider generating PT-AEs against a black-box target model via twisting the pitch feature of environmental sounds.

\subsection{Two-stage Black-box Attack Formulation}
We now formulate the black-box PT-AE attack strategy against a target speaker in a target speaker recognition model. The attack strategy consists of two stages.

\begin{figure}[t]
    \centering
    \includegraphics[width=0.48\textwidth]{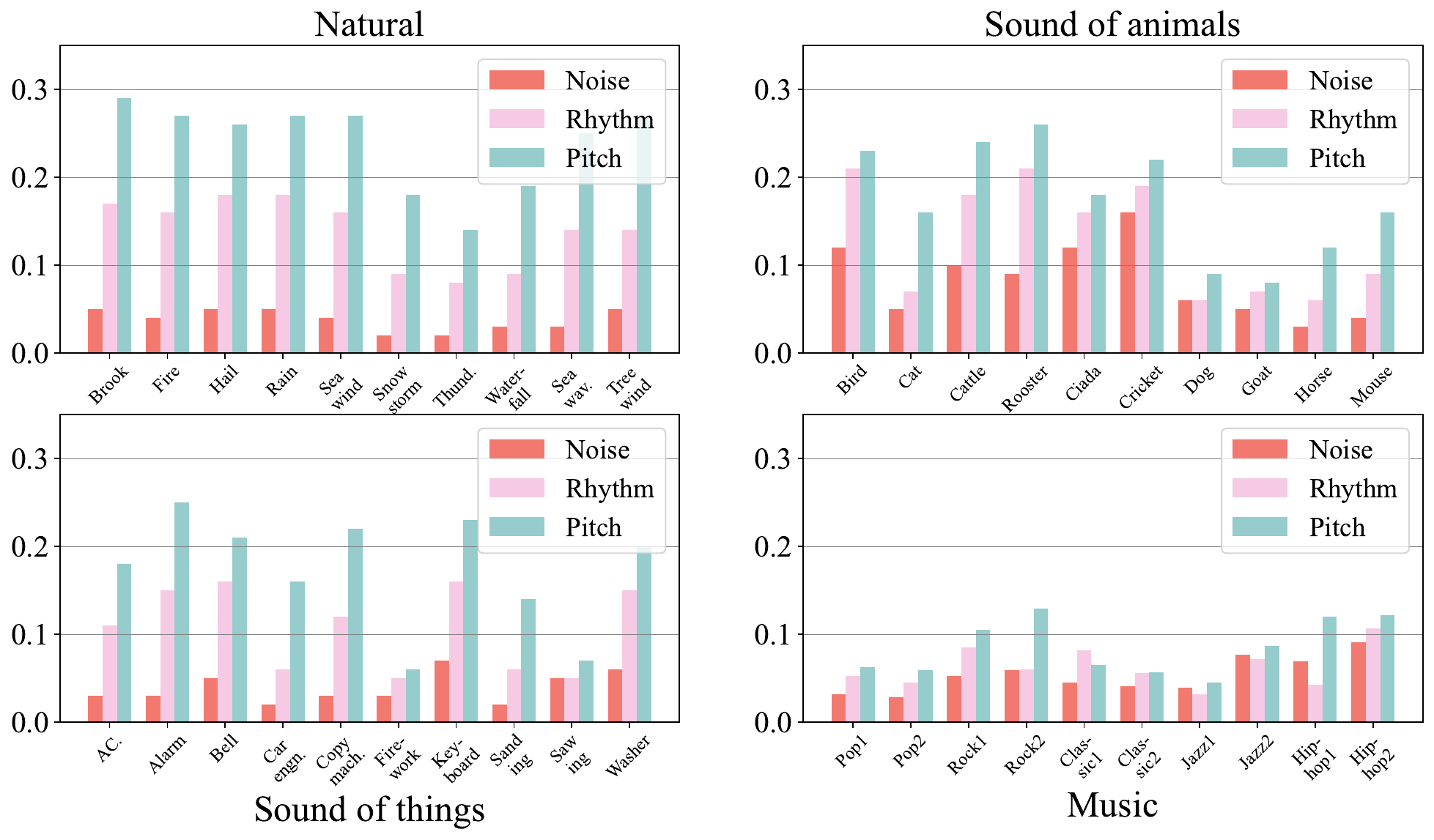}
    \caption{TPR of different optimized carriers.}
    \label{Fig:twisted_feature_TPR}
    \vspace{-0.5cm}
\end{figure}

In the first stage, the attacker needs to determine a set of candidate environmental sounds as there are a wide range of environmental sounds available and not all of them can be effective against the target speaker (as shown in Figure.~\ref{Fig:twisted_feature_TPR}). To this end, we first build a PT-surrogate model for the attacker, evaluate the TPR of each type of environmental sounds based on the surrogate model, and choose $K$ sounds with the best TPRs to form the candidate set. Then, we pre-process each environmental sound in the candidate set by shifting its pitch to obtain its best TPR, and obtain a new candidate set of $K$ pitch-shifted sounds, denoted by $\{\delta_k\}_{k\in [1, K]}$.

In the second stage, we build additional PT-surrogate models for the attacker. We use the same parrot speech samples generated for the target speaker and speech samples of different other speakers to build each PT model. Denote all $N$ PT-surrogate models as $\{\mathcal J_n\}_{n\in [1, N]}$. We employ an ensemble-based method \cite{dong2018boosting,gao2020patch,lin2019nesterov,liu2016delving,wang2021enhancing,xie2019improving}, which linearly combines the loss functions of all the surrogate models (i.e., the ensemble loss), to further improve the transferability of PT-AEs. The attack can be formulated as finding the optimal carrier weights $\gamma_k$ for the pitch-twisted candidate set $\{\delta_k\}_{k\in [1, K]}$ to minimize the ensemble loss:
\begin{eqnarray}
\text{Objective:} && \!\!\!\!\!\!\! \arg\min_{\gamma_k}  \Sigma_{n=1}^N w_n\mathcal{J}_n \left(x+\Sigma_{k=1}^K\gamma_k\delta_k,y_t \right) + \nonumber\\
&& ~~~~~~~~~~~ c\, \text{SRS} \! \left(x,x+\Sigma_{k=1}^K\gamma_k\delta_k \right) \label{Eq:loss}\\
\text{Subject to:} && \!\!\!\!\!\!\! \Sigma_{k=1}^K \gamma_k \leq \epsilon \label{Eq:norm}
\end{eqnarray}
where $x$ is the original speech to be perturbed to generate the attack speech; $y_t$ is the target speaker's label; \eqref{Eq:norm} limits the total energy of the AE carrier within the threshold $\epsilon$; and we uniformly set the model weights $w_n = 1/N$. The optimization \eqref{Eq:loss} is a problem to find multiple carrier weights $\{\gamma_k\}$  with a non-differentiable objective function (because of the perception metric of SRS), we adopt the simultaneous perturbation stochastic approximation (SPSA), which employs a gradient estimation method to optimize the large-scale unknown parameters, to solve \eqref{Eq:loss}. 
We set the uniform weight of each surrogate model \cite{liu2016delving}. To ensure the loss of each surrogate model is in the same range, we convert the cross-entropy loss into a probability via the softmax function. In this way, the loss of each model is in the range of $[0,1]$.

\begin{table*}[t]
    \centering
    \caption{The evaluation of different attacks in digital line.}
    \begin{adjustbox}{center,max width=0.70\linewidth}
    \begin{tabular}{c|cccccccc|cccccccc|cccccccc|cc}
    \toprule
\multicolumn{27}{c}{Intra-gender}                                                                                                                                                                                                                                                                                                                                                                                                                                                                                                                                                                                                                                                                                                                                                                                                                                                                                                                                                                          \\ \midrule
Tasks   & \multicolumn{8}{c|}{CSI}                                                                                                                                                                                                                                                                                        & \multicolumn{8}{c|}{OSI}                                                                                                                                                                                                                                                                                        & \multicolumn{8}{c}{SV}                                                                                                                                                                                                                                                                                                                                        \\ \midrule
Models  & \multicolumn{2}{c|}{\begin{tabular}[c]{@{}c@{}}Deep\\ Speaker\end{tabular}} & \multicolumn{2}{c|}{\begin{tabular}[c]{@{}c@{}}ECAPA-\\ TDNN\end{tabular}} & \multicolumn{2}{c|}{\begin{tabular}[c]{@{}c@{}}GMM-\\ UBM\end{tabular}} & \multicolumn{2}{c|}{\begin{tabular}[c]{@{}c@{}}i-vector-\\ PLDA\end{tabular}} & \multicolumn{2}{c|}{\begin{tabular}[c]{@{}c@{}}Deep\\ Speaker\end{tabular}} & \multicolumn{2}{c|}{\begin{tabular}[c]{@{}c@{}}ECAPA-\\ TDNN\end{tabular}} & \multicolumn{2}{c|}{\begin{tabular}[c]{@{}c@{}}GMM-\\ UBM\end{tabular}} & \multicolumn{2}{c|}{\begin{tabular}[c]{@{}c@{}}i-vector-\\ PLDA\end{tabular}} & \multicolumn{2}{c|}{\begin{tabular}[c]{@{}c@{}}Deep\\ Speaker\end{tabular}} & \multicolumn{2}{c|}{\begin{tabular}[c]{@{}c@{}}ECAPA-\\ TDNN\end{tabular}} & \multicolumn{2}{c|}{\begin{tabular}[c]{@{}c@{}}GMM-\\ UBM\end{tabular}} & \multicolumn{2}{c|}{\begin{tabular}[c]{@{}c@{}}i-vector-\\ PLDA\end{tabular}} & \multicolumn{2}{c}{\textbf{Average}}          \\ \midrule
Metrics & ASR                                   & SRS                                & ASR                                   & SRS                               & ASR                                  & SRS                             & ASR                                    & SRS                                 & ASR                                   & SRS                                & ASR                                   & SRS                               & ASR                                 & SRS                              & ASR                                    & SRS                                 & ASR                                   & SRS                                & ASR                                   & SRS                               & ASR                                 & SRS                              & ASR                                    & SRS                                 & \textbf{ASR}                   & \textbf{SRS} \\
FakeBob & 25.8\%                                & 2.9                                & 26.7\%                                & 3.6                               & 10.6\%                              & 3.2                             & 29.2\%                                 & 3.0                                 & 4.2\%                                 & 2.9                                & 5.8\%                                 & 3.1                               & 6.7\%                               & 3.2                              & 9.2\%                                  & 3.1                                 & 3.0\%                                 & 2.8                                & 5.8\%                                 & 2.6                               & 8.3\%                               & 2.7                              & 5.8\%                                  & 3.2                                 & \textbf{11.3\%}                & \textbf{2.9} \\
Occam   & 45.8\%                                & 2.1                                & 41.7\%                                & 2.1                               & 46.7\%                               & 2.2                             & 47.5\%                                 & 2.4                                 & 5.0\%                                 & 1.6                                & 5.8\%                                 & 1.9                               & 4.2\%                               & 2.1                              & 2.5\%                                  & 2.4                                 & 5.8\%                                & 2.0                                & 5.8\%                                & 1.9                               & 5.0\%                               & 2.2                              & 4.2\%                                  & 2.1                                 & \textbf{19.2\%}                & \textbf{2.1} \\
Smack   & 74.1\%                                & 3.5                                & 45.8\%                                & 2.3                               & 44.2\%                               & 3.6                             & 48.3\%                                 & 3.3                                 & 10.0\%                                & 3.2                                & 13.3\%                                & 3.6                               & 9.2\%                               & 3.5                              & 8.3\%                                  & 2.6                                 & 12.5\%                                & 3.5                                & 13.3\%                                & 3.4                               & 11.7\%                              & 2.1                              & 9.2\%                                  & 2.6                                 & \textbf{29.9\%}                & \textbf{3.1} \\
QFA2SR  & 76.7\%                                & 2.2                                & 70.8\%                                & 2.4                               & 76.7\%                               & 2.1                             & 77.5\%                                 & 2.1                                 & 26.7\%                                & 2.8                                & 31.7\%                                & 2.3                               & 28.3\%                              & 1.9                              & 30.0\%                                 & 2.1                                 & 30.8\%                                & 2.3                                & 29.2\%                                & 1.9                               & 32.5\%                              & 2.6                              & 28.3\%                                 & 2.5                                 & \textbf{40.0\%}                & \textbf{2.3} \\
PT-AEs    & 80.8\%                                & 4.8                                & 79.2\%                                & 4.4                               & 78.3\%                               & 4.3                             & 75.0\%                                 & 4.3                                 & 54.2\%                                & 4.2                                & 56.7\%                                & 3.7                               & 52.5\%                              & 4.4                              & 57.5\%                                 & 3.9                                 & 55.0\%                                & 3.9                                & 56.7\%                                & 3.4                               & 54.2\%                              & 4.1                              & 50.8\%                                 & 4.2                                 & \textbf{60.2\%}                & \textbf{4.1} \\ \midrule
\multicolumn{27}{c}{Inter-gender}                                                                                                                                                                                                                                                                                                                                                                                                                                                                                                                                                                                                                                                                                                                                                                                                                                                                                                                                                                          \\ \midrule
Metrics & ASR                                   & SRS                                & ASR                                   & SRS                               & ASR                                  & SRS                             & ASR                                    & SRS                                 & ASR                                   & SRS                                & ASR                                   & SRS                               & ASR                                 & SRS                              & ASR                                    & SRS                                 & ASR                                   & SRS                                & ASR                                   & SRS                               & ASR                                 & SRS                              & ASR                                    & SRS                                 & \textbf{ASR}                   & \textbf{SRS} \\
FakeBob & 17.5\%                                & 2.9                                & 18.3\%                                & 3.6                               & 13.3\%                               & 3.0                             & 12.5\%                                 & 2.3                                 & 2.5\%                                 & 2.9                                & 1.7\%                                 & 2.7                               & 4.2\%                               & 2.6                              & 2.5\%                                  & 2.4                                 & 2.5\%                                 & 2.1                                & 1.7\%                                 & 2.8                               & 3.3\%                               & 2.7                              & 2.5\%                                  & 2.9                                 & \textbf{6.9\%}                 & \textbf{2.8} \\
Occam   & 26.7\%                                & 3.2                                & 25.8\%                                & 2.6                               & 23.3\%                               & 2.5                             & 21.7\%                                 & 2.1                                 & 5.8\%                                 & 2.8                                & 10.0\%                                & 2.1                               & 10.8\%                              & 2.3                              & 7.5\%                                  & 1.5                                 & 11.7\%                                & 3.0                                & 9.2\%                                 & 2.7                               & 10.0\%                              & 2.6                              & 6.7\%                                  & 2.2                                 & \textbf{14.1\%}                & \textbf{2.6} \\
Smack   & 21.7\%                                & 3.4                                & 26.7\%                                & 3.4                               & 19.2\%                               & 3.0                             & 17.5\%                                 & 3.6                                 & 12.5\%                                & 3.2                                & 14.2\%                                & 3.1                               & 13.3\%                              & 2.8                              & 15.8\%                                 & 2.7                                 & 11.7\%                                & 3.3                                & 15.8\%                                & 3.1                               & 14.2\%                              & 2.9                              & 15.0\%                                 & 2.7                                 & \textbf{16.9\%} & \textbf{3.2} \\
QFA2SR  & 46.7\%                                & 1.9                                & 35.8\%                                & 2.2                               & 43.3\%                               & 2.6                             & 35.8\%                                 & 2.4                                 & 21.7\%                                & 1.5                                & 24.2\%                                & 1.6                               & 25.8\%                              & 2.6                              & 27.5\%                                 & 2.8                                 & 26.7\%                                & 2.1                                & 23.3\%                                & 2.3                               & 26.7\%                              & 2.4                              & 27.5\%                                 & 2.2                                 & \textbf{29.7\%}                & \textbf{2.3} \\
PT-AEs    & 71.7\%                                & 4.3                                & 70.8\%                                & 4.3                               & 70.0\%                               & 4.6                             & 66.7\%                                 & 5.1                                 & 45.8\%                                & 3.8                                & 48.3\%                                & 3.6                               & 46.7\%                              & 3.5                              & 49.1\%                                 & 3.7                                 & 46.7\%                                & 3.9                                & 48.3\%                                & 3.6                               & 49.1\%                              & 3.8                              & 48.3\%                                 & 4.1                                 & \textbf{54.6\%}                & \textbf{3.9}  \\ \bottomrule
    \end{tabular}
    \end{adjustbox}
    \label{Tab:digital_line}
    \vspace{-0.5cm}
    \end{table*}

\section{Experimental Evaluations}\label{Sec:evaluation}
In this section, we measure the impacts of our PT-AE attack in real-world settings. We first describe our setups and then present and discuss experimental results.

\subsection{Experimental Settings}\label{sec:experimental_settings}
\noindent\textbf{The settings of the PT-AE attack:}  We select 3 CNN and 3 TDNN models to build $N=6$ PT models with different parameters for ensembling in \eqref{Eq:loss}. Each PT model has the same one-sentence knowledge (8-second speech) of the target speaker, which is selected from the LibriSpeech \cite{panayotov2015librispeech} or VoxCeleb1 \cite{nagrani2017voxceleb} datasets. We randomly choose 6-16 speakers from the VCTK dataset as other speakers to build each PT model. We choose $K=50$ carriers from the 200 environmental sound carriers in \cite{gemmeke2017audio} to form the candidate set for the attacker and can shift the pitch of a sound up/down by up to 25 semitones. The total energy threshold $\epsilon$ is set to be 0.08.

{\hlb \noindent\textbf{Computational cost:}
We observe that the ensemble loss in \eqref{Eq:loss} typically converges after 500 steps of updating the carrier weights. However, we find that, like gradient descent, SPSA might not always reach the optimal solution and can get stuck in a local minimum. In addition, the presence of a large number of carrier weights can intensify this issue. To address it, we adopt the strategy from \cite{carlini2017towards}, and randomly initialize the weights of carriers $\gamma_k$ 50 times. We then select the carrier weights with the minimal ensemble loss to enhance the transferability of PT-AEs. The maximum computational cost during generating one PT-AE is 25,000 search steps. 
}

\noindent\textbf{Target speaker recognition systems:} We aim to evaluate the attacks against two major types of speaker recognition systems: i) digital-line evaluations: we directly forward AEs to the open-source systems in the digital audio file format (16-bit PCM WAV) to evaluate the attack impact. ii) over-the-air evaluations: we perform over-the-air attack injections to the real-world smart devices.

\noindent\textbf{Evaluation metrics:}
(i) Attack effectiveness: we use attack success rate (ASR) to evaluate the percentage of AEs that can be successfully recognized as the target speaker in a speaker recognition system. (ii) Perception quality: we evaluate the perception quality of an AE via the metric of SRS.

\subsection{Evaluations of Digital-line Attacks}\label{sec:digital-line}
\noindent\textbf{Digital-line setups:}
We consider choosing 4 different target models from statistical-based, i.e., GMM-UBM and i-vector-PLDA \cite{Kaldi}, and DNN-based, i.e., DeepSpeaker \cite{li2017deep} and ECAPA-TDNN \cite{desplanques2020ecapa} models. To increase the diversity of target models, we aim to choose 3 males and 3 females from LibriSpeech and VoxCeleb1. For each gender, we randomly select 1 or 2 speakers from LibriSpeech then randomly select the other(s) from VoxCeleb1. We choose around 15-second speech from each speaker to enroll with each speaker recognition model. {\hlb The performance of each target speaker recognition model is shown in Appendix~\ref{sec:speaker_recognition}.}

\noindent\textbf{Results of digital-line attacks:}
In digital-line evaluations, we measure the performance of each attack strategy by generating 240 AEs (40 AEs for each target speaker) against each target speaker recognition model. We separate the results by the intra-gender (i.e., the original speaker whose speech is used for AE generation is the same-gender as the target speaker) and inter-gender scenario (the original and target speakers are not the same-gender, indicating more distinct speech features). We also evaluate the attacks against three tasks: CSI, OSI, and SV.

Table~\ref{Tab:digital_line} shows the ASRs and SRSs of AEs generated by our PT-AE attack strategy, compared with other attack strategies, against CSI, OSI, and SV tasks. It is noted from Table~\ref{Tab:digital_line} that in the intra-gender scenario, the PT-AE attack and QFA2SR (e.g., 60.2\% for PT-AE attack and 40.0\% for QFA2SR) can achieve higher averaged ASRs (over all three tasks) than other attacks (e.g., 11.3\% for FakeBob, 19.2\% for Occam, and 29.9\% Smack). At the same time, the results of averaged SRS reveal that the perception quality of the PT-AE attack (e.g., 4.1 for PT-AE attack and 3.1 for Smack) is better than other attacks (e.g., 2.3 for QFA2SR, 2.1 for Occam, and 2.9 for FakeBob). 
In addition, it can be observed that in the inter-gender scenario, the ASRs and SRSs become generally worse. For example, the ASR of FakeBob changes from 11.3\% to 6.9\% from the intra-gender to inter-gender scenario. But we can see that our PT-AE attack is still effective in terms of both average ASR (e.g., 54.6\% for PT-AE attack vs 29.7\% for QFA2SR) and average SRS (e.g., 3.9 for PT-AE attack vs 3.2 for Smack). The results in Table~\ref{Tab:digital_line} demonstrate that the PT-AE attack is the most effective in achieving both black-box attack success and perceptual quality.

\subsection{Impacts of Attack Knowledge Levels}\label{sec:knowledge} 
\noindent\textbf{\hlb{1) Impacts of speech length on attack effectiveness:}} By default, we build each PT model in our attack using an 8-second speech sample from the target speaker. We are interested in how the attacker's knowledge affects the PT-AE effectiveness. We assume that the attacker knows the target speaker's speech from 2 to 16 seconds and constructs different PT models based on this varying knowledge to create PT-AEs.

\begin{figure}[!t]
    \centering
    \includegraphics[width=0.478\textwidth]{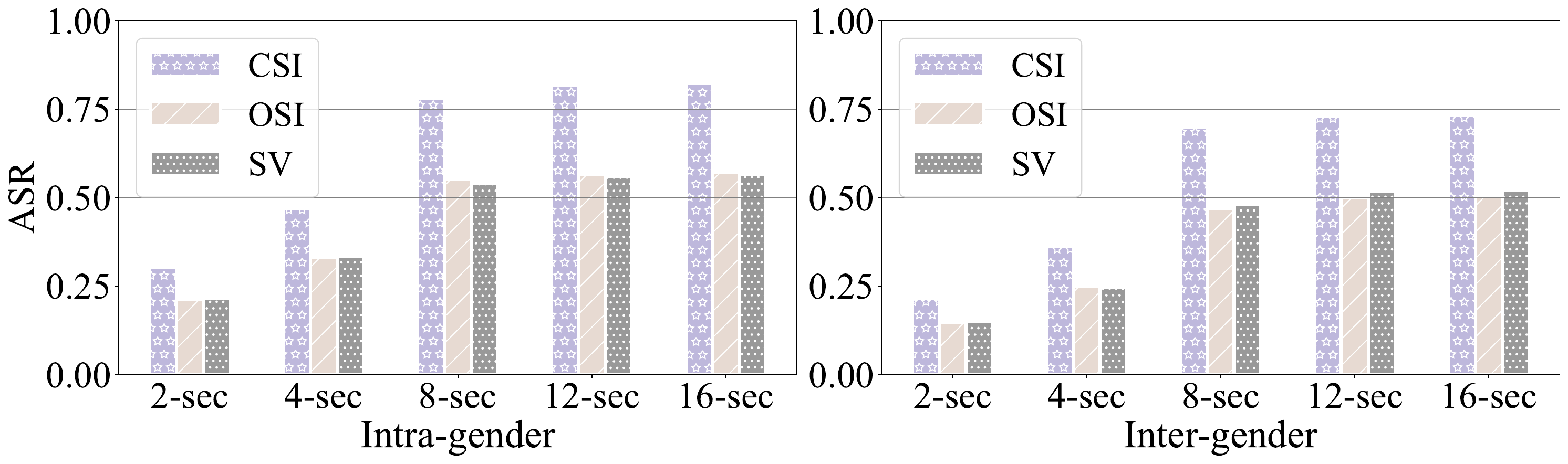}
    \caption{Evaluation on different attack knowledge levels.}
    
    \label{Fig:knowledge}
    \vspace{-0.8cm}
\end{figure}

\noindent\textbf{Results analysis:} Fig.~\ref{Fig:knowledge} shows the ASRs of PT-AEs under different knowledge levels. We can see that more knowledge can increase the attacker's ASR. When the attack knowledge starts to increase from 2 to 8 seconds, the ASR increases substantially (e.g., 21.3\% to 55.2\% against OSI in the intra-gender scenario). When it continues to increase to 16 seconds, the ASR exhibits a slight increase. 
{\hlb
One potential explanation is that the ASR can be influenced by the differences in the architecture and training data between the surrogate and target models. Meanwhile, the one-shot VC method could also reach a performance bottleneck in converting parrot samples using even longer speech. In addition, increasing the speech length does not always indicate the increase of phoneme diversity, which can be also important in speech evaluation \cite{Mines1978Frequency,Brundage1989Measurement}. Existing studies \cite{liu2021any,wang2021vqmivc} highlighted that phonemes represent an important feature of the voiceprint to train the VC model. Thus, we aim to explore further how phoneme diversity (in addition to sentence length) can influence the ASR.
}

{\hlb
\noindent\textbf{2) Impacts of phoneme diversity:}
Since there is no clear, uniform definition for phoneme diversity in previous VC studies \cite{liu2021any,wang2021vqmivc}, we define it as the number of unique phonemes present in a given speech segment. It is worth noting that while some phonemes might appear multiple times in the segment, each is counted only once towards phoneme diversity. This approach is taken because, from an attacker's perspective, unique phonemes are more valuable than repeated ones. While unique phonemes contribute distinct voiceprint features to a VC model, repeated phonemes, can be easily replicated and offer less distinctiveness \cite{liu2021any}.

To evaluate the impact of phoneme diversity on ASR, we choose speech samples of target speakers that have different phoneme diversities but are of the same length (measured by seconds). From our observations in existing datasets (e.g., LibriSpeech), a shorter speech sample can exhibit a higher phoneme diversity than a longer speech sample. This allows us to select speech samples with significantly different levels of phoneme diversity under the same speech length constraint.

We establish low and high phoneme diversity groups in speech segments of the same length to better understand the impact of phoneme diversity on attack effectiveness. In particular, for each level of speech length (e.g., 8-second) in a dataset, we first rank the speech sample of each target speaker by phoneme diversity, then group the top half of all samples (with high values of phoneme diversity) as the high phoneme diversity group and the bottom half as the low diversity group. In this way, the low phoneme diversity group has fewer distinctive phonemes than the high group, offering enough difference regarding attack knowledge for comparison.

We construct our attack knowledge speech set using the speech samples of 3 male and 3 female speakers from LibriSpeech and VoxCeleb1, consistent with the digital-line setups detailed in Section~\ref{sec:digital-line}.  Our goal is to capture various phoneme diversities under different speech lengths. Table~\ref{Tab:phoneme_detail} shows the average phoneme diversity and the total number of phonemes of speech samples in the low and high diversity groups under the same level of speech length (2 to 16 seconds). Table~\ref{Tab:phoneme_detail} demonstrates that the phoneme diversity increases as the speech length increases. Moreover, we find that the phoneme diversity can vary evidently even when the number of total phonemes is similar. For the 8-second category, the low phoneme diversity group has an average diversity of 18.6, while the high diversity group has 24.2. Despite this difference, they have a similar total number of phonemes (80.4 vs 80.6). 

Then, under each level of speech length (2, 4, 8, 12, 16 seconds) for each target speaker (3 male and 3 female speakers), we use speech samples from the low and high phoneme diversity groups for parrot training and generate 90 PT-AEs from each group. This resulted in a total of 5,400 PT-AEs for the phoneme diversity evaluation. 
\begin{table}[t]
    \centering
    \caption{{\hlb Phoneme diversities with different speech lengths.}}
    \begin{adjustbox}{center,max width=0.67\linewidth}
        \begin{threeparttable}
        \begin{tabular}{c|cc|cc|cc|cc|cc}
            \midrule
                           & \multicolumn{2}{c|}{2-second} & \multicolumn{2}{c|}{4-second} & \multicolumn{2}{c|}{8-second} & \multicolumn{2}{c|}{12-second} & \multicolumn{2}{c}{16-second} \\ \midrule
            Averaged       & Diversity       & Total       & Diversity       & Total       & Diversity       & Total       & Diversity        & Total       & Diversity        & Total       \\ \midrule
            Low-diversity  & 5.4             & 12.4        & 10.2            & 23.0        & 18.6            & 80.4        & 26.4             & 100.8       & 32.2             & 134.8       \\ \midrule
            High-diversity & 6.4             & 13.2        & 14.6            & 23.4        & 24.2            & 80.6        & 31.4             & 102.0       & 37.4             & 139.4       \\ \midrule
            \end{tabular}
            \begin{tablenotes}
                \footnotesize
                \item  'Diversity' and 'Total' indicate the phoneme diversity and the number of total phonemes, respectively. 'Low-diversity' and 'High-diversity' indicate the groups with low and high phoneme diversities, respectively.
            \end{tablenotes}
        \end{threeparttable}
        \end{adjustbox}
        \label{Tab:phoneme_detail}
        \vspace{-0.1cm}
        \end{table}

}

\begin{figure}[t]
        \centering
        \includegraphics[width=0.48\textwidth]{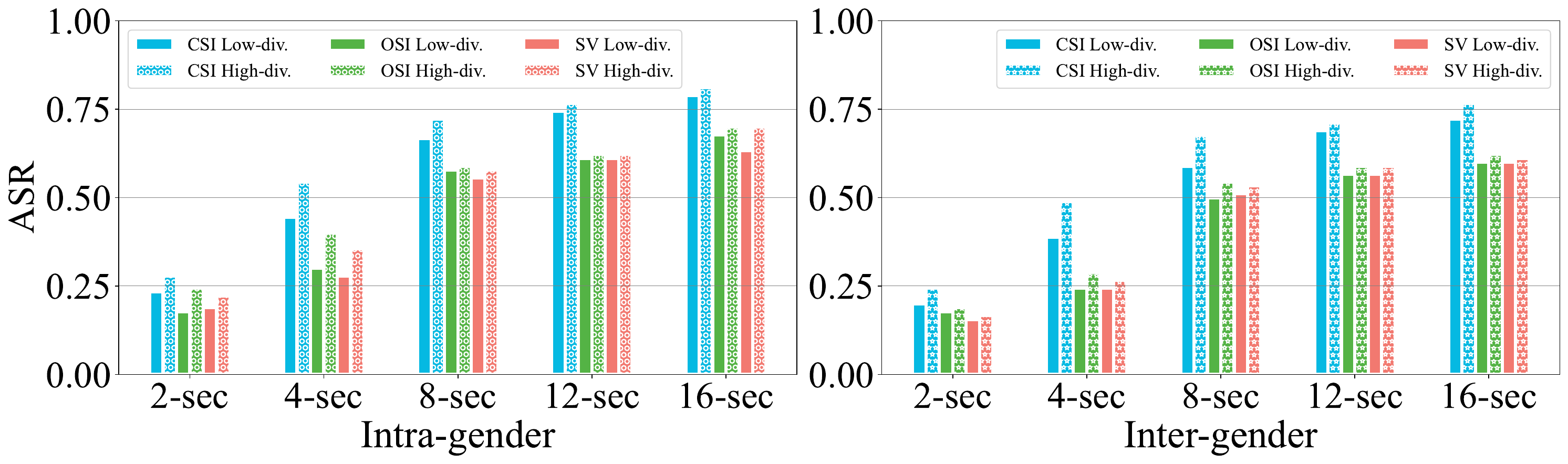}
        \caption{{\hlb Evaluation on phoneme diversity.}}
        \label{Fig:phoneme_diversity}
        \vspace{-0.7cm}
    \end{figure}

{\hlb


\textbf{Results analysis:} Fig.~\ref{Fig:phoneme_diversity} shows the ASRs of PT-AEs generated from low and high diversity groups against CSI, OSI, and SV tasks. It can be seen from the figure that the high-diversity group-based PT-AEs have a higher ASR than the low-diversity ones in both intra-gender and inter-gender scenarios. For example, the inter-gender ASRs are 47.70\% (low-diversity) vs 55.56\% (high-diversity). The largest difference in ASR is observed in the 4-second case in the CSI task for the intra-gender scenario, with a maximum difference of 10.0\%. The results show that using speech samples with high phoneme diversity for parrot training can indeed improve the attack effectiveness of PT-AEs. 

In addition, we calculate via Pearson's coefficients \cite{hauke2011comparison} the correlation of the ASR with each of the methods to measure the attack knowledge level, including measuring the speech length, counting the total number of phonemes, and using the phoneme diversity. We find that phoneme diversity achieves the highest Pearson's coefficient of 0.9692 in comparison with using speech length (0.9341) and counting the total number of phonemes (0.9574). As a result, the phoneme diversity for measuring the attack knowledge is the most related to the attack effectiveness, while using the speech length or the total number of phonemes can still be considered adequate as they both have high Pearson's coefficients. 
}

\begin{table}[t]
    \centering
    \caption{Experimental results on smart devices.}
    \begin{adjustbox}{center,max width=0.685\linewidth}
    \begin{tabular}{c|c|cc|cc|cc|cc|cc}
    \toprule
    \multicolumn{12}{c}{Intra-gender}                                                                                                                                                                                                \\ \midrule
    \multirow{2}{*}{\begin{tabular}[c]{@{}c@{}}Smart\\ Devices\end{tabular}} & Methods & \multicolumn{2}{c|}{FakeBob} & \multicolumn{2}{c|}{Occam} & \multicolumn{2}{c|}{Smack} & \multicolumn{2}{c|}{QFA2SR} & \multicolumn{2}{c}{PT-AEs} \\ \cmidrule{2-12}
                                                                             & Tasks   & ASR           & SRS         & ASR          & SRS        & ASR           & SRS       & ASR           & SRS        & ASR          & SRS       \\ \midrule
    Amazon Echo                                                              & OSI     & 0/12          & N/A         & 1/12         & 1.89       & 2/12          & 4.45      & 3/12          & 2.60       & 7/12         & 4.33      \\ \midrule
    Amazon Echo                                                              & SV      & 0/12          & N/A         & 2/12         & 2.01       & 2/12          & 4.53      & 4/12          & 2.72       & 7/12         & 5.08      \\ \midrule
    Google Home                                                              & SV      & 0/12          & N/A         & 0/12         & N/A        & 1/12          & 3.96      & 3/12          & 2.55       & 5/12         & 4.49      \\ \midrule
    Apple HomePod                                                               & SV      & 2/12          & 2.15        & 3/12         & 3.16       & 3/12          & 5.09      & 5/12          & 3.12       & 9/12         & 5.16      \\ \midrule
    \textbf{Average}                                                         & \textbf{-} & \textbf{4.2\%} & \textbf{2.15} & \textbf{12.5\%} & \textbf{2.35} & \textbf{16.7\%}  & \textbf{4.51} & \textbf{31.3\%}  & \textbf{2.75} & \textbf{58.3\%} & \textbf{4.77}      \\ \midrule
    \multicolumn{12}{c}{Inter-gender}                                                                                                                                                                                                \\ \midrule
    \multicolumn{1}{c|}{}                                                     & \multicolumn{1}{c|}{Tasks} & ASR            & SRS           & ASR             & SRS           & ASR              & SRS           & ASR              & SRS           & ASR             & SRS           \\ \midrule
    Amazon Echo                                                              & OSI     & 0/12          & N/A         & 1/12         & 1.26       & 2/12          & 3.89      & 2/12          & 2.27       & 5/12         & 4.15      \\ \midrule
    Amazon Echo                                                              & SV      & 0/12          & N/A         & 1/12         & 1.35       & 1/12          & 4.12      & 3/12          & 2.03       & 6/12         & 4.27      \\ \midrule
    Google Home                                                              & SV      & 0/12          & N/A         & 0/12         & N/A        & 1/12          & 3.11      & 2/12          & 1.92       & 4/12         & 4.53      \\ \midrule
    Apple HomePod                                                               & SV      & 1/12          & 1.59        & 2/12         & 2.59       & 2/12          & 4.14      & 4/12          & 3.10       & 8/12         & 4.86      \\ \midrule
    \textbf{Average}                                                         & \textbf{-} & \textbf{2.1\%} & \textbf{1.59} & \textbf{8.3\%}  & \textbf{1.73} & \textbf{12.5\%} & \textbf{3.82} & \textbf{22.9\%} & \textbf{2.33} & \textbf{47.9\%} & \textbf{4.45}      \\ \bottomrule
    \end{tabular}
    \end{adjustbox}
    \label{Tab:smart_device}
    \vspace{-0.4cm}
\end{table}

\subsection{Evaluations of Over-the-air Attacks}\label{sec:over_the_air}
Next, we focus on attacking the smart devices in the over-the-air scenario. We consider three popular smart devices: Amazon Echo Plus \cite{Amazon_example}, Google Home Mini\cite{Google}, and Apple HomePod (Siri) \cite{Applesiri}. For speaker enrollment, we use 3 male and 3 female speakers from Google's text-to-speech platform to generate the enrollment speech for each device. We only use an 8-second speech from each target speaker to build our PT models. We consider OSI and SV tasks on Amazon Echo, and the SV task on Apple HomePod and Google Home. Similarly, we evaluate the different attacks in both intra-gender and inter-gender scenarios. For each attack strategy, we generate and play 24 AEs using a JBL Clip3 speaker to each smart device with a distance of 0.5 meters.

\noindent\textbf{Results analysis:} Table~\ref{Tab:smart_device} compares different attack methods against the smart devices under various tasks. We can see that our PT-AE attack can achieve average ASRs of 58.3\% (intra-gender) and 47.9\% (inter-gender) and at the same time the average SRSs of 4.77 (intra-gender) and 4.45 (inter-gender). By contrast, QFA2SR has the second-best ASRs of 31.3\% (intra-gender) and 22.92\% (inter-gender); however, it has a substantially lower perception quality compared with the PT-AE attack and Smack, e.g., 2.75 (QFA2SR) vs 4.51 (Smack) vs 4.77 (PT-AE attack) in the intra-gender scenario. We also find that FakeBob and Occam appear to be ineffective with over-the-air injection as zero ASR is observed against Amazon Echo and Google Home. Overall, the over-the-air results demonstrate that the PT-AEs generated by the PT-AE attack can achieve a high ASR with good perceptual quality.
{\hlb
Additionally, we also evaluated the robustness of PT-AEs over distance, the results can be found in Table~\ref{Tab:distance} in Appendix~\ref{sec:distance}.
}

{\hlb\subsection{Contribution of Each Component to ASR}\label{sec:contribution}
As the PT-AE generation involves three major design components, including parrot training, choosing carriers, and ensemble learning, to enhance the overall transferability, we propose to evaluate the contribution of each individual component to the ASR. Our methodology is similar to the One-at-a-time (OAT) strategy in \cite{duan2022perception}. Specifically, we remove and replace each design component with an alternative, baseline approach (as a baseline attack), while maintaining the other settings the same in generating PT-AEs, and then compare the resultant ASR with the ASR of no-removing PT-AEs (i.e., the PT-AEs generated without removing/replacing any design component). Through this method, we can determine how each component contributes to the overall attack effectiveness. 

We use the same over-the-air attack setup as described in Section~\ref{sec:over_the_air}. For each baseline attack, we craft 96 AEs for both intra and inter-gender scenarios. These AEs are played on each smart device by the same speaker at the same distance. We present the experimental setup and results regarding evaluating the contribution of each design component as follows.

\begin{table}[t]
    \centering
    \caption{{\hlb ASRs with removing each design component.}}
    \begin{adjustbox}{center,max width=0.70\linewidth}
        \begin{tabular}{c|c|ccccc}
            \toprule
                                                                                         &            & Amazon-OSI & Amazon-SV & Google-SV & Apple-SV & \textbf{Average} \\ \midrule
            {\bf No removing}                                                                & PT-AEs          & 50.0\%    & 54.2\%    & 37.5\%    & 70.8\%   & \textbf{53.1\%}  \\ \midrule
            \textbf{1) No PT}                                                              & Non-PT AEs          & 29.2\%     & 33.3\%    & 25.0\%    & 37.5\%   & \textbf{31.3\%}  \\ \midrule
            \multirow{2}{*}{\textbf{\begin{tabular}[c]{@{}c@{}}2) No environ- \\ mental sound\end{tabular}}}                                                    & Noise           & 25.0\%     & 33.3\%    & 25.0\%    & 33.3\%   & \textbf{29.2\%}  \\ \cmidrule(lr){2-7} 
                                                                                         & Featute-twisted & 33.3\%     & 37.5\%    & 25.0\%    & 37.5\%   & \textbf{33.3\%}  \\ \midrule
            \multirow{4}{*}{\begin{tabular}[c]{@{}c@{}}\textbf{3) No or insufficient }\\ \textbf{ensemble learning}\end{tabular}} & Single PT-CNN   & 29.2\%     & 33.3\%    & 20.8\%    & 41.7\%   & \textbf{31.3\%}  \\ \cmidrule(lr){2-7} 
                                                                                         & Single PT-TDNN  & 29.2\%     & 37.5\%    & 20.8\%    & 41.7\%   & \textbf{32.3\%}  \\ \cmidrule(lr){2-7} 
                                                                                         & Multiple PT-CNN         & 41.7\%     & 45.8\%    & 29.2\%    & 58.3\%   & \textbf{43.8\%}  \\ \cmidrule(lr){2-7} 
                                                                                         & Multiple PT-TDNN        & 45.8\%     & 45.8\%    & 33.3\%    & 58.3\%   & \textbf{45.8\%}  \\ \bottomrule
            \end{tabular}
    \end{adjustbox}
    \label{Tab:contribution}
    \vspace{-0.4cm}
    \end{table}

\noindent\textbf{1) Parrot training:} Rather than training the surrogate models with parrot speech, we directly use the target speaker's one-sentence (8-second) speech for enrollment with the surrogate models. These surrogate models, which we refer to as non-parrot-training (non-PT) models, are trained on the datasets that exclude the target speakers' speech samples. 

\noindent\textbf{Results:} As shown in Table~\ref{Tab:contribution} (the ``No PT'' row), we observe a significant ASR difference between non-PT-based AEs and no-removing PT-AEs. For example, in the Amazon-SV task, PT-AEs achieve an ASR of 54.2\%, which is 20.9\% higher than the 33.3\% ASR of non-PT AEs. Overall, the average ASR for PT-AEs is 21.8\% higher than that of non-PT AEs. This substantial performance gap is primarily filled by adopting parrot training.

\noindent\textbf{2) Environmental sound carrier:} To understand the contribution of the feature-twisted environment sound carrier, we use two baseline attacks related to noise and feature-twisted carriers. i) Noise carriers, we employ the PGD attack to generate the AEs based on the PT models through ensemble learning, setting $\epsilon =0.05 $ to control the $L_{\infty}$ norm. ii) Feature-twisted carriers, as discussed in Section~\ref{sec:combining_carriers}, we shift the pitch of the original speech up or down by up to 25 semitones to create a pitch-twisted set. We use this set to solve the problem in \eqref{Eq:loss} via finding the optimal weights for the twisted-pitch carriers, with a total energy threshold of $\epsilon = 0.08 $.

\noindent\textbf{Results:} Table~\ref{Tab:contribution} (the ``no environmental sound'' rows) indicates that environmental-sound-based PT-AEs hold a distinct advantage over other carriers in terms of attack effectiveness. We note that when we exclude the feature-twisted environmental sound carriers and rely solely on either the noise or feature-twisted carriers, the average ASR drops by 23.9\% (vs. noise carrier) and 19.8\% (vs. feature-twisted carrier). These findings show that utilizing feature-twisted environmental sounds can significantly enhance the attack effectiveness.

\noindent\textbf{3) Ensemble learning:} We note that our ensemble-based model in \eqref{Eq:loss} combines multiple CNN and TDNN models. To evaluate the contribution of ensemble learning, we design two sets of experiments. First, we replace the ensemble-based model in \eqref{Eq:loss} with just a single PT-CNN or PT-TDNN model to compare the ASRs. Second, we replace \eqref{Eq:loss} with an ensemble-based model, which only consists of multiple (in particular 6 in experiments) surrogate models under the same CNN or TDNN architecture (i.e., no ensembling across different architectures). 

\noindent\textbf{Results:} We can observe in Table~\ref{Tab:contribution} (the ``no or insufficient ensemble learning'' rows) that the single PT-CNN and PT-TDNN models only have average ASRs of 31.3\% and 32.3\%, respectively. If we do adopt ensemble learning but combine surrogate models under the same architecture, the average ASRs can be improved to 43.8\% and 45.8\% under multiple PT-CNN and PT-TDNN models, respectively. By contrast, no-removing PT-AEs achieve the highest average ASR of 53.1\%. 



}

{\hlb
In summary, the three key design components for PT-AEs, i.e., parrot training, feature-twisted environmental sounds, and ensemble learning, improve the average ASR by 21.8\%, 21.9\%, and 21.3\%, respectively, when compared with their individual baseline replacements. As a result, they are all important towards the black-box attack and have approximately equal contribution to the overall ASR.
}

\begin{figure}[t]
    \centering
    \includegraphics[width=0.485\textwidth]{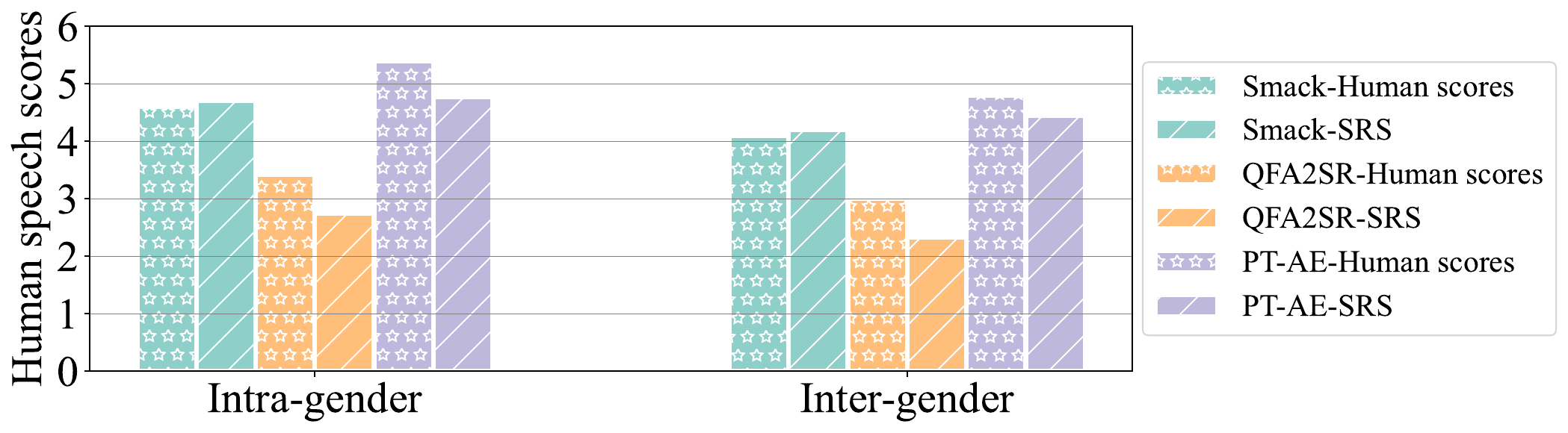}
    \caption{Human evaluation on the AEs.}
    \label{Fig:human_AEs}
    \vspace{-0.8cm}
\end{figure}

\subsection{Human Study of AEs Generated in Experiments}\label{sec:human_PTAEs}
We have used the metric of SRS based on regression prediction built upon the human study in Section~\ref{Sec:Human_study} to assess that the PT-AEs have better perceptual quality than AEs generated by other attack methods in experimental evaluations. We now conduct a new round of human study to see whether PT-AEs generated in the experiments are indeed rated better than other AEs by human participants. {\hlb Specifically, we have recruited} additional 45 student volunteers (22 females and 23 males), with ages ranging from 18 to 35. They are all first-time participants and have no knowledge of the previous human study in Section~\ref{Sec:Human_study}. Following the same procedure, we ask each volunteer to rate each pair of original and PT-AE samples.

Fig.~\ref{Fig:human_AEs} shows the average human speech scores of Smack, QFA2SR, and our attack. {\hlb We can see that} PT-AEs generated by our attack are rated higher than Smack and QFA2SR. In the intra-gender scenario, the average human score of our attack is 5.39, which is higher than Smack (4.61) and QFA2SR (3.62). {\hlb The score for each method drops slightly in the inter-gender scenario. The results align with the SRS findings in Table~\ref{Tab:smart_device}. We also find SRS scores are close to human scores. In the inter-gender scenario, SRS predicts our PT-AEs perceptual quality as 4.45, close to the human average of 4.8.} The results of Fig.~\ref{Fig:human_AEs} further validates that the PT-AEs have better perceptual quality than AEs generated by other methods.

\subsection{Discussions}
\noindent\textbf{Ethical concerns and responsible disclosure:} Our smart device experiments did not involve any person's private information. All the experiments were set up in our local lab. 
{\hlb We have reported our findings to manufacturers (Amazon, Apple, and Google). All manufacturers thanked our research and disclosure efforts aimed at safeguarding their services. Google responded promptly to our investigations, confirming that there is a voice mismatch issue and closed the case as they stated that the attack requires the addition of a malicious node. We are still in communication with Amazon and Apple.}

{\hlb
We also discuss potential defense strategies against PT-AEs. Due to the page limit, we have presented the defense discussion in Appendix~\ref{Sec:defense}.}

\section{Related work}\label{Sec:related}
\noindent\textbf{White-box attacks:}
Adversarial audio attacks \cite{carlini2018audio, yuan2018commandersong, li2020advpulse, taori2019targeted, wang2020towards, chen2020devil, du2020sirenattack, zheng2021black,du2020sirenattack, chen2019real, zheng2021black} can be categorized into white-box and black-box attacks depending on their attack knowledge level. White-box attacks \cite{carlini2018audio, qin2019imperceptible} assumed the knowledge of the target model and leveraged the gradient information of the target model to generate highly effective AEs. Some recent studies aimed at improving the practicality of white-box attacks \cite{li2020advpulse,guo2022specpatch} via adding the perturbation to the original speech signal without synchronization, albeit still assuming nearly full knowledge of the target model.

\noindent\textbf{Query-based black-box attacks:} Existing black-box attacks \cite{chen2019real,zheng2021black,taori2019targeted,wang2020towards,liu2022evil,yusmack} assumed no access to the internal knowledge of target models, and most black-box attacks attempted to know the target model via a querying (or probing) strategy. The query-based attacks \cite{chen2019real,du2020sirenattack, zheng2021black,yusmack,liu2022evil} needed to interact with the target model to get the internal prediction scores \cite{chen2019real,wang2020towards,chen2020devil,yusmack} or hard label results \cite{zheng2021black,liu2022evil}. A large number of queries were necessary for the black-box attack to be effective. For example, Occam \cite{zheng2021black} needed over 10,000 queries to achieve a high ASR. This makes the attack strategy cumbersome to launch, especially in over-the-air scenarios. The PT-AE attack does not require any probing to the target model.

\noindent\textbf{Transfer-based black-box attacks:} The transfer-based attacks \cite{abdullah2019hear,duan2022perception,chen2023qfa2sr} commonly assumed no interaction or limited probing \cite{chen2020devil} to the target model. For example, Kenansville \cite{abdullah2019hear} manipulated the phoneme of the speech to achieve an untargeted attack.
QFA2SR \cite{chen2023qfa2sr} focused on building the surrogate models with specific ensemble strategies to enhance the transferability of AEs by assuming knowing several speech samples of all the enrolled speakers of the target model. Compared with QFA2SR, we further minimize the knowledge and only assume a short speech sample of the target speaker for the attacker. Even with the most limited attack knowledge, we propose a new PT-AE strategy that creates more effective AEs against the target model.

\noindent\textbf{Audio attacks considering the perception quality:} Some recent studies  \cite{qin2019imperceptible,guo2022specpatch,liu2022evil} leveraged the psychoacoustic feature to optimize the carriers and improve the perception of AEs. Meanwhile, \cite{duan2022perception, yusmack} manipulated the features of an audio signal to create AEs with good perceptual quality. In addition, there are audio attack strategies \cite{zhang2017dolphinattack, carlini2016hidden,abdullah2019practical, yuan2018commandersong} focusing on improving the stealthiness of the AEs. For example, dolphin attack \cite{zhang2017dolphinattack} used ultrasounds to generate imperceptible AEs. The human study in this work defines the metric of SRS to quantify the speech quality using a similar regression procedure motivated by the qDev model in \cite{duan2022perception} that was created to measure the music quality. We then design a new TPR framework built upon the SRS metric to jointly evaluate both the transferability and perception of PT-AEs.

\section{Conclusion}\label{Sec:conclusion}
In this work, we investigated using the minimum knowledge of a target speaker's speech to attack a black-box target speaker recognition model. We extensively evaluated the feasibility of using state-of-the-art VC methods to generate parrot speech samples to build a PT-surrogate model and the generation methods of PT-AEs. It is shown that PT-AEs can effectively transfer to a black-box target model and the proposed PT-AE attack has achieved higher ASRs and better perceptual quality than existing methods against both digital-line speaker recognition models and commercial smart devices in over-the-air scenarios.

\bibliographystyle{plain}
\bibliography{reference}

\appendix

{\hlb

\subsection{Speaker Recognition Models}\label{sec:SR}
\subsubsection{Speaker Recognition Mechanisms} 
Speaker recognition models\cite{TencentVPR,Kaldi,nidadavolu2019investigation,lee2019coral+} are typically categorized into statistical models, such as Gaussian-Mixture-Model (GMM) based Universal Background Model (UBM) \cite{reynolds2000speaker} and i-vector probabilistic linear discriminant analysis (PLDA) \cite{dehak2010front,nandwana2019analysis}, and deep neural network (DNN) models \cite{li2017deep,desplanques2020ecapa}. There are three phases in speaker recognition.
\begin{enumerate}
    \item In the training phase, one key component is to extract the acoustic features of speakers, which are commonly represented by the encoded low-dimensional speech features, (e.g., i-vectors \cite{dehak2010front} and X-vectors \cite{snyder2018x}). Then, these features can be trained by a classifier (e.g., PLDA \cite{ioffe2006probabilistic}) to recognize different speakers.
    \item During the enrollment phase, to make the classifier learn a speaker's voice pattern, the speaker usually needs to deliver several text-dependent (e.g., Siri \cite{Applesiri} and Amazon Echo \cite{Alexa}) or text-independent speech samples to the speaker recognition system. Depending on the number of enrolled speakers, speaker recognition tasks \cite{chen2019real, zheng2021black, yusmack} can be (i) multiple-speaker-based speaker identification (SI) or (ii) single-speaker-based speaker verification (SV).
    \item In the recognition phase, the speaker recognition model will predict the speaker's label or output a rejection result based on the similarity threshold. Specifically, SI can be divided into close-set identification (CSI) and open-set identification (OSI) \cite{deng2022fencesitter, chen2019real}. The former predicts the speaker's label with the highest similarity score, and the latter only outputs a prediction when the similarity score is above the similarity threshold or gives a rejection decision otherwise. SV only focuses on identifying one specific speaker. If the similarity exceeds a predetermined similarity threshold, SV returns an accepted decision. Otherwise, it will return a rejection decision.
\end{enumerate}

\subsubsection{Speaker Recognition Formulations}
Let $y_i$ denote the $i$-th speaker enrolled in group set $\mathcal{Y}$, where $\mathcal{Y} = \{y_1,y_2,\cdots,y_i \}$. Let $S(x,y_i)$ represent the similarity score function which takes the test speech signal $x$ as the input and outputs the similarity score based on the enrolled speaker $y_i \in \mathcal{Y}$.

\noindent\textbf{$\bullet$ CSI:} The CSI task assumes the test speech $x$ always belongs to a speaker in $\mathcal{Y}$, and there is no outsider speaking. The classification function of CSI $f_{_\text{CSI}}(x)$ will output the speaker's label with the highest similarity score, i.e.,
\begin{equation*}
f_{_\text{CSI}}(x)=\arg\max_{y_i \in \mathcal{Y}} S(x,y_i).
\end{equation*}

\noindent\textbf{$\bullet$ OSI:} Different from the CSI task, OSI is able to judge whether the test speech $x$ belongs to $\mathcal{Y}$ or not. And its classification function $f_{_\text{OSI}}(x)$ only outputs a speaker's label when the highest similarity score exceeds the threshold $\theta$.
\begin{equation*}
f_{_\text{OSI}}(x) = \left\{
    \begin{array}{ll}
        \underset{y_i \in \mathcal{Y}}{\arg\max}~S(x,y_i),  & \quad\text{if }  \underset{y_i \in \mathcal{Y}}{\operatorname{max}} \, S(x,y_i) \geq \theta_{_\text{OSI}}, \\
        \text{Reject,} & \quad \text{otherwise},
    \end{array}
\right.
\end{equation*}
where $\theta_{_\text{OSI}}$ is the similarity threshold to reject in OSI.

\noindent\textbf{$\bullet$ SV:} The enrollment set of SV is only one speaker $y_1$ but not multiple speakers, and it also requires the similarity score greater than the threshold.
\begin{equation*}
f_{_\text{SV}}(x) = \left\{
    \begin{array}{ll}
        \text{Accept,}  & \quad\text{if}~S(x,y_1) \geq \theta_{_\text{SV}}, \\
        \text{Reject,} & \quad \text{otherwise},
    \end{array}
\right.
\end{equation*}
where $\theta_{_\text{SV}}$ is the threshold to accept or reject in SV.

\subsection{Comparison of PT and GT Models}\label{sec:PTvsGT}
\noindent\textbf{Constructing PT models:} There are multiple ways to set up and compare PT and GT models. We set up the models based on our black-box attack scenario, in which the attacker knows that the target speaker is trained in a speaker recognition model but does not know other speakers in the model. We first build a GT model using multiple speakers' speech samples, including the target speaker's. To build a PT model for the attacker, we start from the only information that the attacker is assumed to know (i.e., a short speech sample of the target speaker), and use it to generate different parrot speech samples. Then, we use these parrot samples, along with speech samples from a small set of speakers (different from the ones used in the GT model) in an open-source dataset, to build a PT model.

We use CNN and TDNN to build two GT models, called CNN-GT and TDNN-GT, respectively. Each GT model is trained with 6 speakers (labeled from 1 to 6) from LibriSpeech (90 speech samples for training and 30 samples for testing for each speaker). We build 6 CNN-based PT models, called CNN-PT-$i$, and 6 TDNN-based PT models, called TDNN-PT-$i$, where $i$ ranges from 1 to 6 and indicates that the attacker's targets speaker~$i$ in the GT model and uses only one of his/her speech samples to generate parrot samples, which are used together with samples from other 3 to 8 speakers randomly selected from VCTK (none is in the GT models), to train a PT model. 

\noindent\textbf{Evaluation metrics:}
We aim to compare the 12 PT models with the 2 GT models when recognizing the attacker's target speaker. Existing studies \cite{li2021modeldiff,korotcov2017comparison} have investigated how to compare different machine learning models via the classification outputs. We follow the common strategy and validate whether PT models have the performance similar to GT models via common classification metrics, including Recall \cite{davis2006relationship}, Precision \cite{ferri2002learning}, and F1-Score \cite{nagrath2021ssdmnv2}, where Recall measures the percentage of correctly predicted target speech samples out of the total actual target samples, Precision measures the proportion of the speech which is predicted as the target label indeed belongs to the target speaker, and F1-Score provides a balanced measure of a model's performance which is the harmonic mean of the Recall and Precision. To test each PT model (targeting speaker~$i$) and measure the output metrics compared with GT models, we use 30 ground-truth speech samples of speaker~$i$ from LibriSpeech and 30 samples of every other speaker from VCTK in the PT model.

\noindent\textbf{Results analysis and discussion:} Fig.~\ref{Fig:PT_VS_GT} shows the classification performance of PT and GT models. It is observed from the figure that CNN-GT/TDNN-GT achieves the highest Recall, Precision, and F1-Score, which range from 0.97 to 0.98. We can also see that most PT models have slightly lower yet similar classification performance as the GT models. For example, CNN-PT-1 has similar performance to TDNN-GT (Recall: 0.93 vs 0.98; Precision: 0.96 vs 0.98; F1-Score 0.95 vs 0.98). The results indicate that a PT model, just built upon one speech sample of the target speaker, can still recognize most speech samples from the target speaker, and also reliably reject to label other speakers as the target speaker at the same time. The worst-performing model TDNN-PT-4 achieves a Recall of 0.82 and a Precision of 0.86, which is still acceptable to recognize the target speaker. Overall, we note that the PT models can achieve similar classification performance compared with the GT models. Based on the findings, we are motivated to use a PT model to approximate a GT model in generating AEs, and aim to further explore whether PT-AEs are effective to transfer to a black-box GT model.

\begin{figure}[t]
    \centering
    \includegraphics[width=0.48\textwidth]{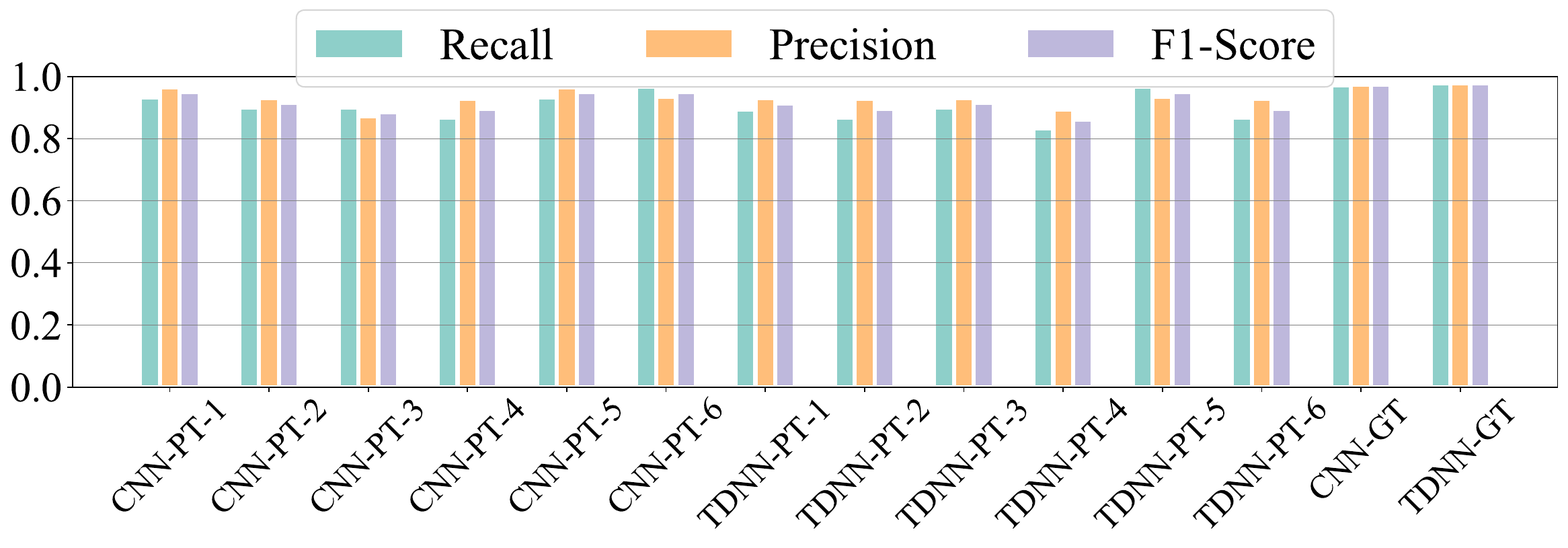}
    \caption{Comparison of PT and GT models.}
    \label{Fig:PT_VS_GT}
    \vspace{-0.6cm}
\end{figure}

}

{\hlb

\subsection{Performance of Digital-line Speaker Recognition Models}\label{sec:speaker_recognition}
Table~\ref{Tab:SR_performance} shows the performance of the target speaker recognition models, where accuracy indicates the percentage of speech samples that are correctly labeled by a model in the CSI task; False Acceptance Rate (FAR) is the percentage of speech samples that belong to unenrolled speakers but are accepted as enrolled speakers; False Rejection Rate (FRR) is the percentage of samples that belong to an enrolled speaker but are rejected; Open-set Identification Error Rate (OSIER) is the equal error rate of OSI-False-Acceptance and OSI-False-Rejection. 

\begin{table}[t]
    \centering
    \caption{Performance of speaker recognition systems.}
    \begin{adjustbox}{center,max width=0.98\linewidth}
    \begin{tabular}{c|c|ccc|cc}
    \toprule
    \multirow{2}{*}{Task} & CSI & \multicolumn{3}{c|}{OSI} & \multicolumn{2}{c}{SV} \\
                           & Accuracy     & FAR          & FRR   & OSIER   & FAR        & FRR       \\ \midrule
    DeepSpeaker            & 98.89\% & 11.42\%      & 1.11\%      & 0.83\%    & 6.96\%    & 0.41\%          \\ \midrule
    ECAPA-TDNN             & 99.58\% & 9.74\%  &0.42\%   & 0.03\%   & 4.87\%       & 0.42\%       \\ \midrule
    GMM-UBM                & 99.44\% & 10.72\% & 5.15\% & 2.65\%  & 10.02\%     & 5.01\%    \\ \midrule
    i-vector-PLDA               & 99.72\% & 7.93\% & 2.36\% & 0.27\%     & 12.25\%      & 0.97\%     \\ \bottomrule
    \end{tabular}
    \end{adjustbox}
    \label{Tab:SR_performance}
    \vspace{-0.2cm}
\end{table}

\subsection{Robustness of PT-AEs over Distance}\label{sec:distance} 

We aim to further evaluate the robustness of the PT-AE attack in the over-the-air scenario with different distances from the attacker to the target. We set different levels of distance between the attacker (i.e., the JBL Clip3 speaker) and a smart device from 0.25 to 4 meters. The results in Table~\ref{Tab:distance} show that the ASR of the PT-AE attack changes over the distance. In particular, we can see that there is no significant degradation of ASR when the distance goes from 0.25 to 0.5 meters as the ASR slightly decreases from 60.4\% to 58.3\% in the inter-gender scenario. There is an evident degradation in ASR when the distance increases from 2.0 to 4.0 meters (e.g., 27.1\% to 14.5\% in the inter-gender scenario). This is due to the energy degradation of PT-AEs when they propagate over the air to the target device. Overall, PT-AEs are quite effective within 2.0 meters given the perturbation energy threshold of $\epsilon=0.08$ set for all experiments.

\begin{table}[t]
\centering
\caption{Evaluation of different distances.}
\begin{adjustbox}{center,max width=0.80\linewidth}
\begin{tabular}{c|c|cccccc}
\toprule
\begin{tabular}[c]{@{}c@{}}Attack\\ Scenarios\end{tabular} & \begin{tabular}[c]{@{}c@{}}Smart\\ Devices\end{tabular} & Distance & 0.25 (m)   & 0.5 (m)    & 1.0 (m)   & 2.0 (m)   & 4.0 (m)   \\ \midrule
\multirow{5}{*}{Intra-gender}                              & Amazon Echo                                             & OSI          & 58.3\% & 58.3\% & 41.7\% & 25.0\% & 16.7\% \\ \cmidrule(lr){2-8}
                                                           & Amazon Echo                                             & SV           & 58.3\% & 58.3\% & 50.0\% & 33.3\% & 16.7\% \\ \cmidrule(lr){2-8}
                                                           & Google Home                                             & SV           & 50.0\% & 41.7\% & 41.7\% & 25.0\% & 16.7\% \\ \cmidrule(lr){2-8}
                                                           & Apple HomePod                                              & SV           & 75.0\% & 75.0\% & 75.0\% & 58.3\% & 33.3\% \\ \cmidrule(lr){2-8}
                                                            & \textbf{Average}                                        & \textbf{-}   & \textbf{60.4\%} & \textbf{58.3\%} & \textbf{52.1\%} & \textbf{35.4\%} & \textbf{20.8\%} \\ \midrule
\multirow{5}{*}{Inter-gender}                              & Amazon Echo                                             & OSI          & 41.7\% & 41.7\% & 25.0\% & 16.7\% & 8.3\%  \\ \cmidrule(lr){2-8}
                                                           & Amazon Echo                                             & SV           & 50.0\% & 50.0\% & 33.3\% & 25.0\% & 16.7\% \\ \cmidrule(lr){2-8}
                                                           & Google Home                                             & SV           & 33.3\% & 33.3\% & 25.0\% & 16.7\% & 8.3\%  \\ \cmidrule(lr){2-8}
                                                           & Apple HomePod                                              & SV           & 66.7\% & 66.7\% & 66.7\% & 50.0\% & 25.0\% \\ \cmidrule(lr){2-8}
                                                           & \textbf{Average}                                        & \textbf{-}   & \textbf{47.9\%} & \textbf{47.9\%} & \textbf{37.5\%} & \textbf{27.1\%} & \textbf{14.5\%} \\ \bottomrule
\end{tabular}
\end{adjustbox}
\label{Tab:distance}
\end{table}

\subsection{Discussion on Defense}\label{Sec:defense}

\noindent\textbf{Potential defense designs:} To combat PT-AEs, there are two major defense directions available: (i) audio signal processing and (ii) adversarial training. Audio signal processing has been proposed to defend against AEs via down-sampling \cite{liu2022evil,zheng2021black}, quantization \cite{yang2018characterizing}, and low-pass filtering \cite{li2020advpulse} to preserve the major frequency components of the original signal while filtering out other components to make AEs ineffective. These signal processing methods may be effective when dealing with the noise carrier \cite{zheng2021black,li2020advpulse,guo2022specpatch}, but are not readily used to filter out PT-AEs based on environment sounds, many of which have similar frequency ranges as human speech. Adversarial training \cite{goodfellow2014explaining, madry2017towards, balaji2019instance, cai2018curriculum, shafahi2019adversarial, tramer2017ensemble, wong2020fast} is one of the most popular methods to combat AEs. The key idea behind adversarial training is to repeatedly re-train a target model using the worst-case AEs to make the model more robust. One essential factor in adversarial training is the algorithm used to generate these AEs for training. For example, recent work \cite{zheng2021black} employed the PGD attack to generate AEs for adversarial training, and the model becomes robust to the noise-carrier-based AEs. {\hlb One potential way for defense is to generate enough AEs that cover a diversity of carriers and varying auditory features for training. Significant designs and evaluations are needed to find optimal algorithms to generate and train AEs to fortify a target model.}

}

\newpage

\end{document}